\documentclass[preprint2]{aastex}

\received{}
\accepted{}
\journalid{}{}
\articleid{}{}
\slugcomment{To appear in {\it the Astrophysical Journal}}

\shortauthors{KOBAYASHI et al.}
\shorttitle{Galactic chemical evolution}

\def\gtsim {>\kern-1.2em\lower1.1ex\hbox{$\sim$}~}   
\def\ltsim {<\kern-1.2em\lower1.1ex\hbox{$\sim$}~}   

\begin{document}

\title{Galactic chemical evolution: Carbon through Zinc}
\author{Chiaki KOBAYASHI$^1$, Hideyuki UMEDA$^2$, Ken'ichi NOMOTO$^2$, Nozomu TOMINAGA$^2$, and Takuya OHKUBO$^2$}
\affil{$^1$ Division of Theoretical Astronomy, National Astronomical Observatory of Japan, Mitaka-shi, Tokyo 181-8588, Japan;
chiaki@th.nao.ac.jp}
\affil{$^2$ Department of Astronomy, School of Science,
University of Tokyo, Bunkyo-ku, Tokyo 113-0033, Japan}

\begin{abstract}
We calculate the evolution of heavy element abundances from C to Zn in the solar neighborhood adopting our new nucleosynthesis yields.
Our yields are calculated for wide ranges of metallicity ($Z=0-Z_\odot$) and the explosion energy (normal supernovae and hypernovae), based on the light curve and spectra fitting of individual supernovae.
The elemental abundance ratios are in good agreement with observations.
Among the $\alpha$-elements, O, Mg, Si, S, and Ca show a plateau at [Fe/H] $\ltsim -1$, while Ti is underabundant overall.
The observed abundance of Zn ([Zn/Fe] $\sim0$) can be explained only by the high energy explosion models, which requires a large contribution of hypernovae.
The observed decrease in the odd-Z elements (Na, Al, and Cu) toward low [Fe/H] is reproduced by the metallicity effect on nucleosynthesis.
The iron-peak elements (Cr, Mn, Co, and Ni) are consistent with the observed mean values at $-2.5 \ltsim$ [Fe/H] $\ltsim -1$, and the observed trend at the lower metallicity can be explained by the energy effect.
We also show the abundance ratios and the metallicity distribution functions of the Galactic bulge, halo, and thick disk. Our results suggest that the formation timescale of the thick disk is $\sim 1-3$ Gyr.
\end{abstract}


\section{Introduction}

Elemental abundance ratios are the treasure house of information on star formation and galaxy formation because different types of supernovae produce different heavy elements with different timescales \citep[e.g.,][]{tin80,pagel1997,matteucci2001}.
High resolution spectroscopy gives the elemental abundance patterns of individual stars in our Galaxy and nearby dwarf spheroidal galaxies, and of high-redshift quasar absorption line systems.
These observations have shown different abundance patterns, which suggest different chemical enrichment histories of these objects and probably different initial mass functions (IMFs) depending on environment.
The re-ionization of the universe at very early epoch as suggested by the WMAP result may require a different population for the first stars \citep[e.g.,][]{cir03}.
However, to discuss these issues, nucleosynthesis yields have involved too many inconsistencies with recent observations.

There exist two distinct types of supernova explosions \citep[e.g.,][]{arnett1996,alex1997}: One is 
Type II supernovae (SNe II), which are the core collapse-induced explosions
of massive stars ($\gtsim \, 8M_\odot$) 
with short lifetimes of $10^{6-7}$ yrs,
and produce more $\alpha$-elements (O, Mg, Si, S, Ca, and Ti) 
relative to Fe with respect to the solar ratios (i.e., [$\alpha$/Fe]
$>0$). The other is
Type Ia supernovae (SNe Ia), which are the thermonuclear explosions 
of accreting white dwarfs (WDs) in close binaries 
and produce mostly Fe and little $\alpha$-elements \citep[e.g.,][]{nom94}.
Although the lifetime of SNe Ia has been estimated to be the order of Gyr from the chemical evolution of our Galaxy (e.g., Yoshii, Tsujimoto \& Nomoto 1996),
the simulations of the SN Ia progenitor systems predict that the lifetime of the majority of SNe Ia is shorter than this
(Hachisu, Kato, \& Nomoto 1996; Kobayashi et al. 1998, hereafter K98).
The metallicity effect of SNe Ia that is also predicted by the simulations can solve this problem (K98).

Following the development of observations of individual supernovae and elemental abundances, theory of stellar nucleosynthesis has been improved from Woosley \& Weaver (1995) and Nomoto et al. (1997a, hereafter N97).
From the light curve and spectral fitting of individual supernovae, the mass of progenitors stars $M$, explosion energy $E$, and produced $^{56}$Ni mass (which decayed to $^{56}$Fe) have been obtained (e.g., Nomoto et al. 2004 for a review).
There exist two distinct types of core-collapse supernovae:
One is normal SNe II (including Ib and Ic), which have the explosion energy of $E_{51} \equiv E/10^{51}$ erg $\sim 1$ and produce little iron relative to $\alpha$-elements.
Some SNe II with $M>25M_\odot$ may have smaller energy ($E_{51}<1$), which are called ``faint SNe'' \citep{tur98}.
The other is hypernovae (HNe), which have more than ten times larger explosion energy ($E_{51}\gtsim10$) and produce a certain amount of iron.
HNe may be related to gamma-ray bursts \citep{gal98,iwa98}.

Umeda \& Nomoto (2002, 2005, hereafter UN05) updated the progenitor star models with a metallicity range of $Z=0-0.02$ (see also Limongi \& Chieffi 2003), and calculated the explosive nucleosynthesis yields for larger energies ($E_{51}>1$).
To meet the observed abundance patterns for extremely metal-poor (EMP) stars \citep{cay04,aoki2004,beers2005}, they introduced the mixing and fallback mechanism.
They then succeed in reproducing the observed abundance patterns of some EMP stars (Umeda \& Nomoto 2003, hereafter UN03; UN05), based on the idea that the interstellar medium is not mixed and the EMP stars are enriched only by a single supernova \citep{aud95}.
The trends of iron-peak element abundance patterns for the EMP stars \citep{mcw95,rya96} have been explained by changing a parameter mass-cut \citep{nak99}, while they were explained with the energy dependence: [(Zn,Co)/Fe] increases and [(Mn,Cr)/Fe] decreases for metal poor stars because of higher energy and a larger amount of swept hydrogen mass (UN05).

In this paper, however, we focus on typical yields of core-collapse supernovae that is responsible for the plateau values of the [$\alpha$/Fe]-[Fe/H] relations, which is important when we discuss the galaxy formation and evolution.
Physical parameters in our model are constrained from observations, i.e., the light curve and spectra of individual supernovae and the abundance patterns of metal poor stars in the solar neighborhood.
There may be a scatter in the observed mass-energy relation and [$\alpha$/Fe]-[Fe/H] relations that arise from inhomogeneity of the explosion and of the mixing of the interstellar medium.
However, at $-2.5 \ltsim$ [Fe/H] $\ltsim -1$, chemical enrichment of the interstellar medium is proceeded with many supernova explosions, and the average contribution of supernovae with different mass is imprinted in the [X/Fe]-[Fe/H] diagrams.
First, we give a table set of nucleosynthesis yields both for SNe II and HNe as functions of progenitor mass and metallicity (\S 2).
We then show the evolution of elemental abundance ratios from carbon to zinc in the solar neighborhood using a one-zone chemical evolution model (\S 3).
In \S 4, we construct models for the bulge, halo, and thick disk of the Milky Way Galaxy to meet their metallicity distribution functions, and show the age-metallicity relations. Showing the different evolution of elemental abundance ratios, we extend our discussion to the formation of the thick disk.
\S 5 gives our conclusions.

\section{Nucleosynthesis Yields}

\subsection{Nucleosynthesis Models}

Using the same calculation method as in Umeda \& Nomoto (2002, 2005), we calculate stellar evolution, explosions, and nucleosynthesis for wider ranges of metallicity and energy.
The details of the calculations are described in Umeda et al. (2000) and Tominaga et al. (2006).
The code is based on the Henyey-type stellar evolution code by Nomoto \& Hashimoto (1988) and Umeda et al. (1999). We start calculations from the zero-age main-sequence through core collapse including metallicity-dependent mass loss \citep{dej88,kud89}.
This code runs a nuclear reaction network by Hix \& Thielemann (1996) for nuclear energy generation and nucleosynthesis, including the neutron capture processes.
We adopt the Schwartzshild criterion for convective stability and diffusive convective mixing by Spruit (1992). In this work, a case for relatively fast mixing, $f_k=0.3$, is adopted.
The $^{12}$C ($\alpha$, $\gamma$) $^{16}$O reaction rate, which is still uncertain (Fowler 1984), is chosen to be $1.3$ times the value given in Caughlan \& Fowler (1988).

For a given progenitor model, if the explosion mechanism is specified (or the procedure for the artificial explosion like Woosley \& Weaver (1995) and Limongi \& Chieffi (2003)), the remnant mass is uniquely determined as a function of the explosion energy. However, we do not specify the explosion mechanism and treat the mixing and fallback with free-parameters, especially because the precise explosion mechanism is unknown for hypernovae.
Since we explode the progenitor model when the central density of $3 \times 10^{10}$ g cm$^{-3}$ is reached without calculating further collapse and bounce, our approach may be regarded as simulating a prompt explosion. 
According to the constraint from the light curve and spectral fitting (\S 1), we set two sequences of mass-energy relation; (1) $E_{51}=1$ for all normal SNe II, and (2) $E_{51}=10$, 10, 20, and 30 for the 20, 25, 30, and $40M_\odot$ HNe, respectively.

Using the pre-supernova models, we carry out 1D hydrodynamical simulations of core-collapse explosions using the piecewise parabolic method \citep{col84} with the $\alpha$-nuclear reaction network for the energy generation rate. Then explosive nucleosynthesis is calculated as a post-processing using a larger reaction network of 300 isotopes.
After the postprocess nucleosynthesis calculations, we obtain the final yields setting the mass-cut for SNe II.
The mass-cuts are chosen to eject $M({\rm Fe})\simeq 0.07M_\odot$,
which is constrained from the light curve and spectra of individual supernovae \citep[e.g., ][]{nom04}.
For HNe, we take account of the mixing-fallback mechanism.
The Rayleigh-Taylor mixing and the amount of fallback both depend on the stellar mass, pre-supernova density structure, explosion energy, asphericity, etc., so that its determination requires extremely high resolution calculations.
We thus determine the parameters involved in the mixing and fallback (UN05) to give [O/Fe] $\simeq 0.5$ according to the constraint from the abundance ratios of EMP stars (see \S \ref{sec:xfe}). 

The neutrino process is not included in our nucleosynthesis, but the number of electrons per nucleon, $Y_{\rm e}$, depends on the neutrino process during explosion \citep[e.g., ][]{lie03,jan03}.
Therefore, $Y_{\rm e}$ in the incomplete Si-burning region is set to be $0.4997$ independent of metallicity, while $Y_{\rm e}$ in the other region is kept constant as pre-supernova model, where $Y_{\rm  e} \sim 0.5$ above the oxygen layer and decreases gradually toward the Fe core (UN05). 
The initial mass-cut is determined to be located at the bottom of the $Y_{\rm e} \simeq 0.5$ layer \citep{tom06}.

Tables 1 and 2 give the resultant nucleosynthesis yields in the ejecta in $M_\odot$ after radioactive decays, for SNe II and HNe as functions of the progenitor mass ($M= 13$, 15, 18, 20, 25, 30, and $40 M_\odot$) and metallicity ($Z= 0$, 0.001, 0.004, and $0.02$).
The mass of the pre-supernova star, $M_{\rm final}$, is larger for lower metallicity because of the metal-dependent stellar winds.
The material inside the mass-cut $M_{\rm cut}$ falls onto the remnant, and the rest ($M_{\rm ejecta}=M_{\rm final}-M_{\rm cut}$) is ejected by the supernova explosion.

\begin{figure*}
\center
\includegraphics[width=16cm]{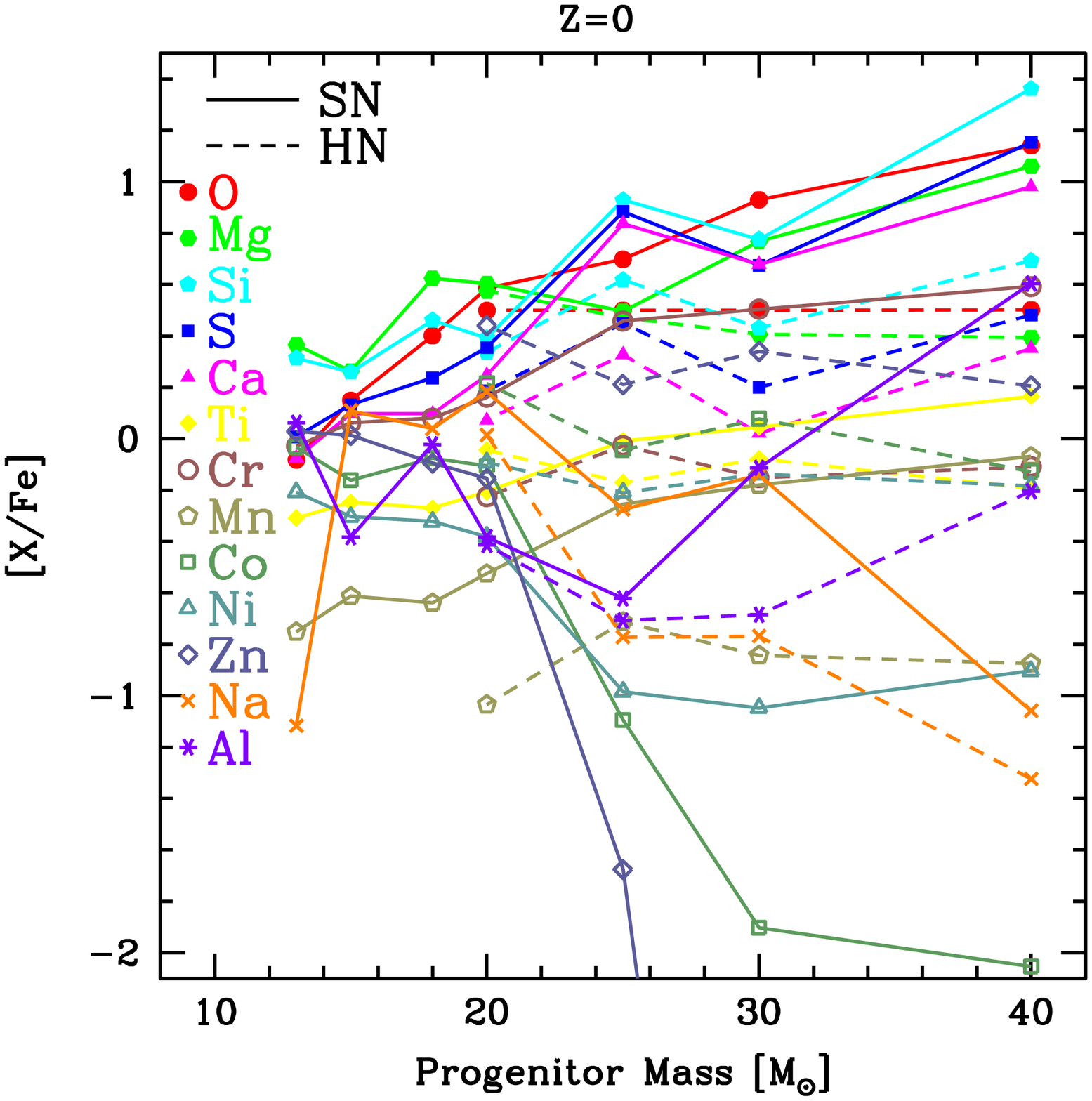}
\caption{\label{fig:yield}
Relative abundance ratios as a function of the progenitor mass with $Z=0$.
The solid and dashed lines show normal SNe II with $E_{51}=1$ and HNe.
}
\end{figure*}

\begin{figure*}
\center
\includegraphics[width=16cm]{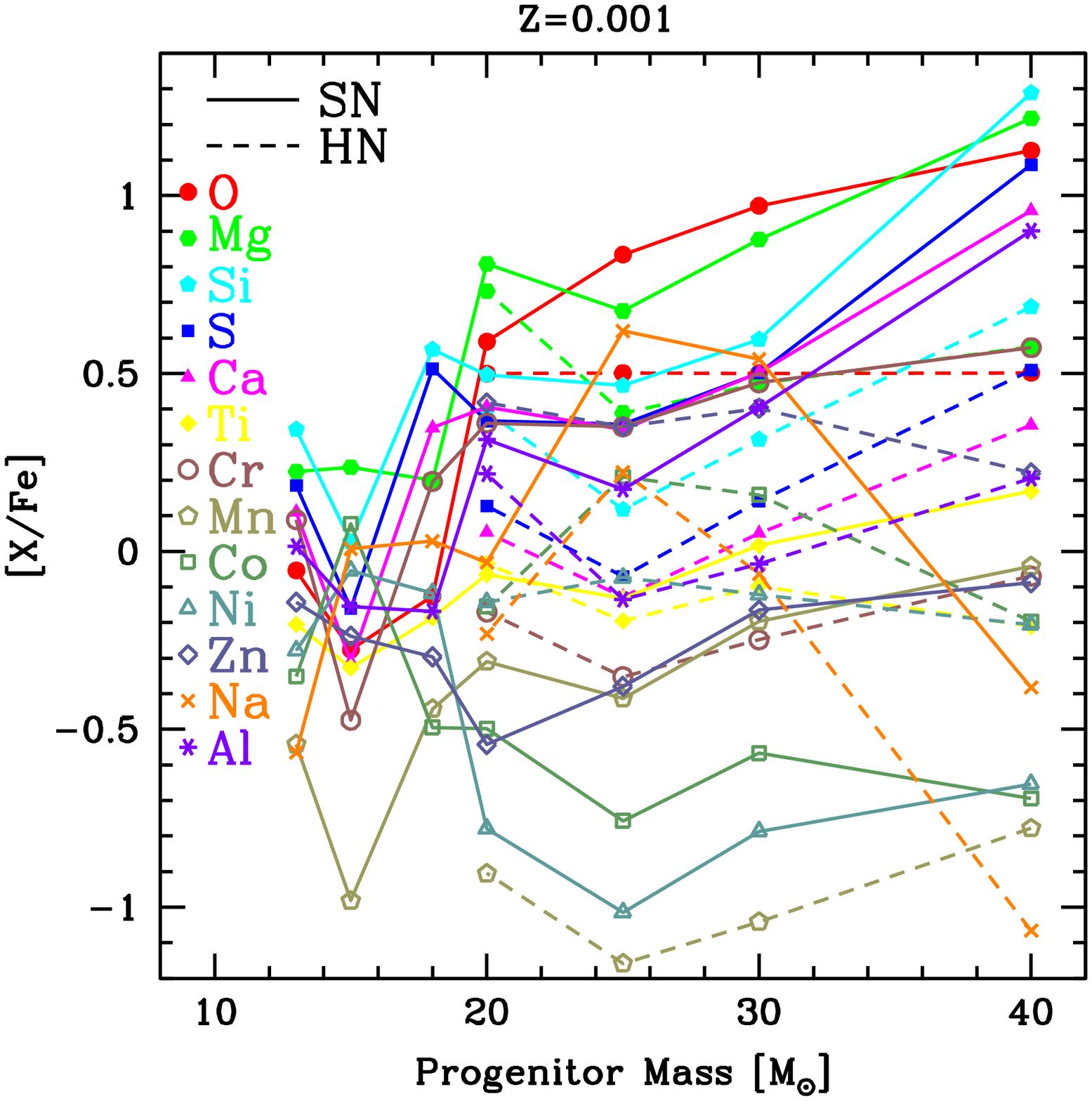}
\caption{\label{fig:yield2}
The same as Fig.\ref{fig:yield} but for $Z=0.001$.
}
\end{figure*}

\begin{figure*}
\center
\includegraphics[width=16cm]{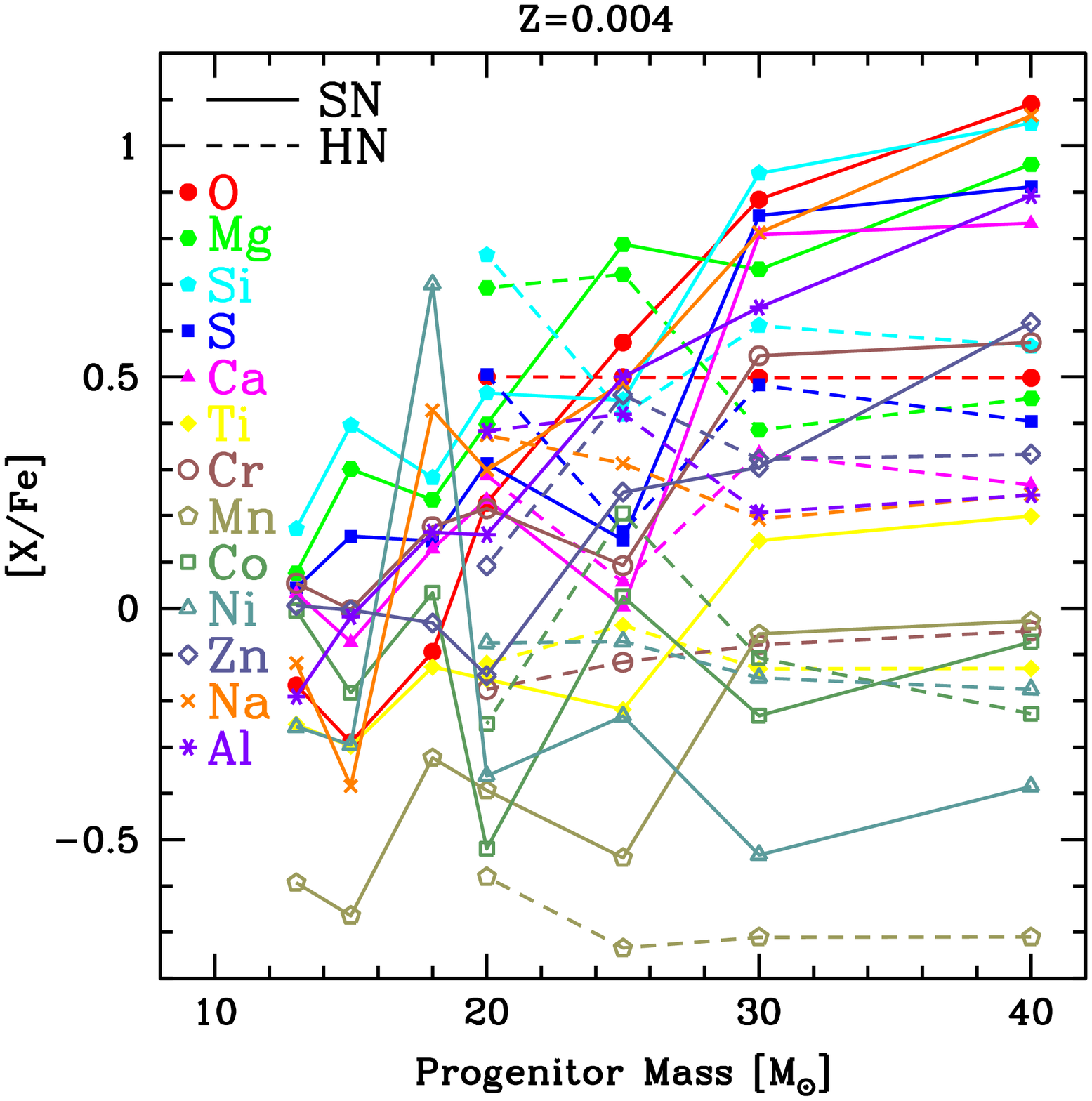}
\caption{\label{fig:yield3}
The same as Fig.\ref{fig:yield} but for $Z=0.004$.
}
\end{figure*}

\begin{figure*}
\center
\includegraphics[width=16cm]{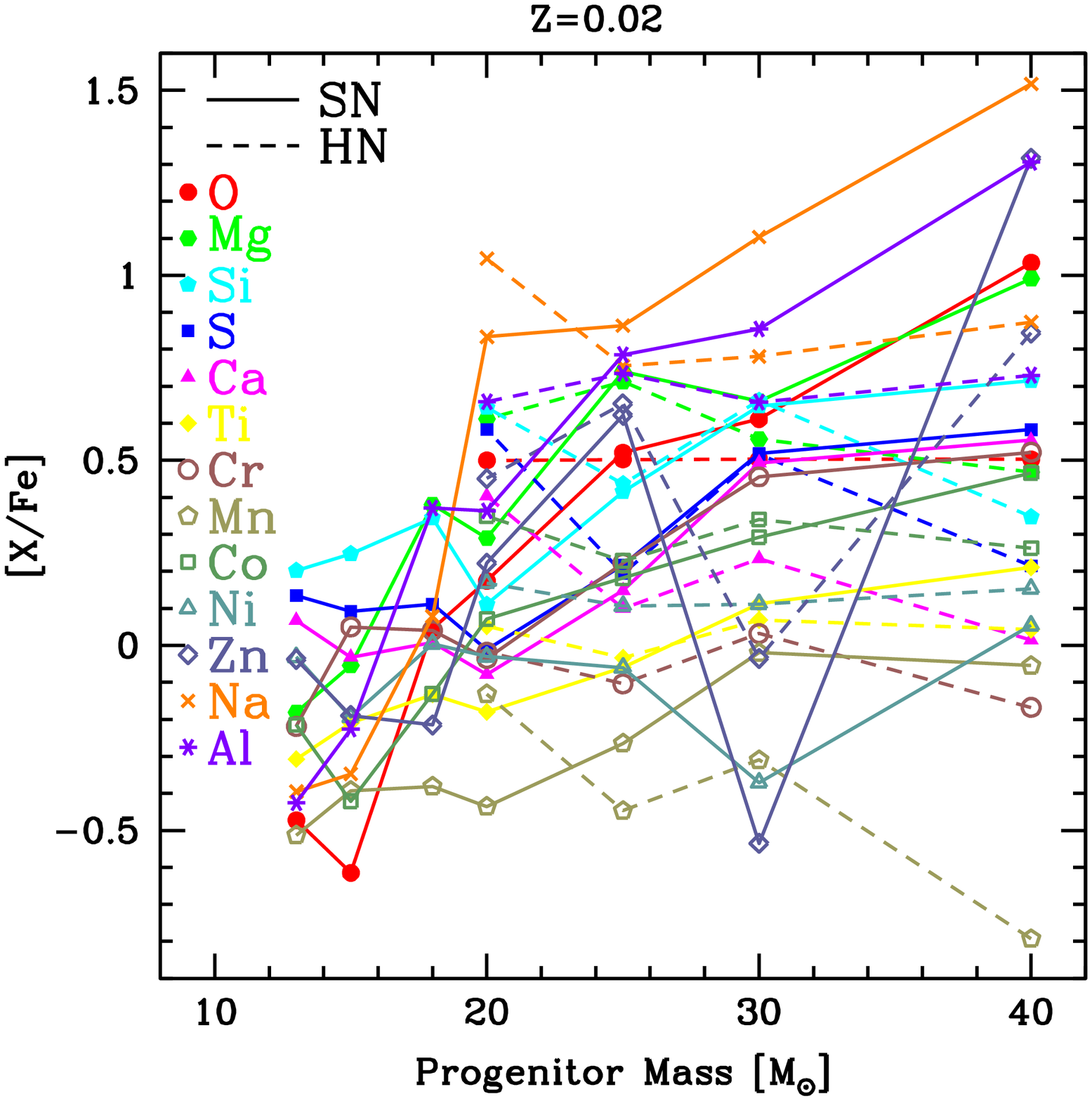}
\caption{\label{fig:yield4}
The same as Fig.\ref{fig:yield} but for $Z=0.02$.
}
\end{figure*}

Figures \ref{fig:yield}-\ref{fig:yield4} show the abundance ratios relative to the solar abundance \citep{and89} in the ejecta as a function of the progenitor mass for given metallicity.
The solid and dashed lines show the SN II and HN yields, respectively.
The yield masses of $\alpha$-elements (O, Ne, Mg, Si, S, Ar, Ca, and Ti) are larger for more massive stars because of the larger mantle mass. 
Since the ejected Fe mass is $\simeq 0.07M_\odot$ for SNe II, being almost independent of the progenitors mass, the abundance ratio [$\alpha$/Fe] is larger for more massive stars.
This mass dependence of [$\alpha$/Fe] is smaller in the present yields than in N97, because N97 adopted a larger ejected Fe mass ($0.15M_\odot$ for $13-15M_\odot$) from the supernova observations around early 1980's, while the present study adopt the smaller Fe mass from the modeling of the recent well-observed supernova light curves.
For HNe, although the Fe mass is larger for more massive stars because of the higher energy, the mass of $\alpha$-elements is also larger, and thus [$\alpha$/Fe] is almost constant independent of the stellar mass.

In the present yields, the abundance ratios of iron-peak elements (Cr, Mn, Co, and Ni) are almost constant with respect to the progenitor mass.
The differences from N97 are due to the new implementation of the mixing-fallback.
Zn and Cu yields are much larger than N97.
This is because the neutron capture processes enhance the neutron-rich Zn and Cu abundances for $Z \ge 0.004$.
For smaller $Z$, the larger Zn production is due to the larger energy and larger electron fraction $Y_{\rm e}$ in the present explosion models.
It might be possible to reduce Zn production for high- or low-mass stars and explain the observed [Zn/Fe] trend by mass sequence, but then the average Zn abundance would be much smaller than observed.
The smaller [Mn/Fe] is also due to large $Y_{\rm e} \sim 0.5$.

\subsection{IMF Weighted Yields}

The mass-energy relation has been obtained from the light curve and spectral fitting for individual supernovae. 
However, there is currently no constraint on the energy distribution function because of the poor statistics.
Therefore, in the chemical evolution model, we should introduce one important parameter to describe the fraction of hypernovae, $\epsilon_{\rm HN}$. 
$\epsilon_{\rm HN}$ may depend on metallicity, and may be constrained by the gamma-ray burst rate.
Here we adopt $\epsilon_{\rm HN}=0.5$ independent of the mass and metallicity.
This gives a good agreement with the [$\alpha$/Fe] plateau against [Fe/H] as shown in \S \ref{sec:xfe}.
Such large $\epsilon_{\rm HN}$ is required from the observed [Zn/Fe] $\sim 0$, especially for low metallicity.

\begin{figure*}
\center
\includegraphics[width=16cm]{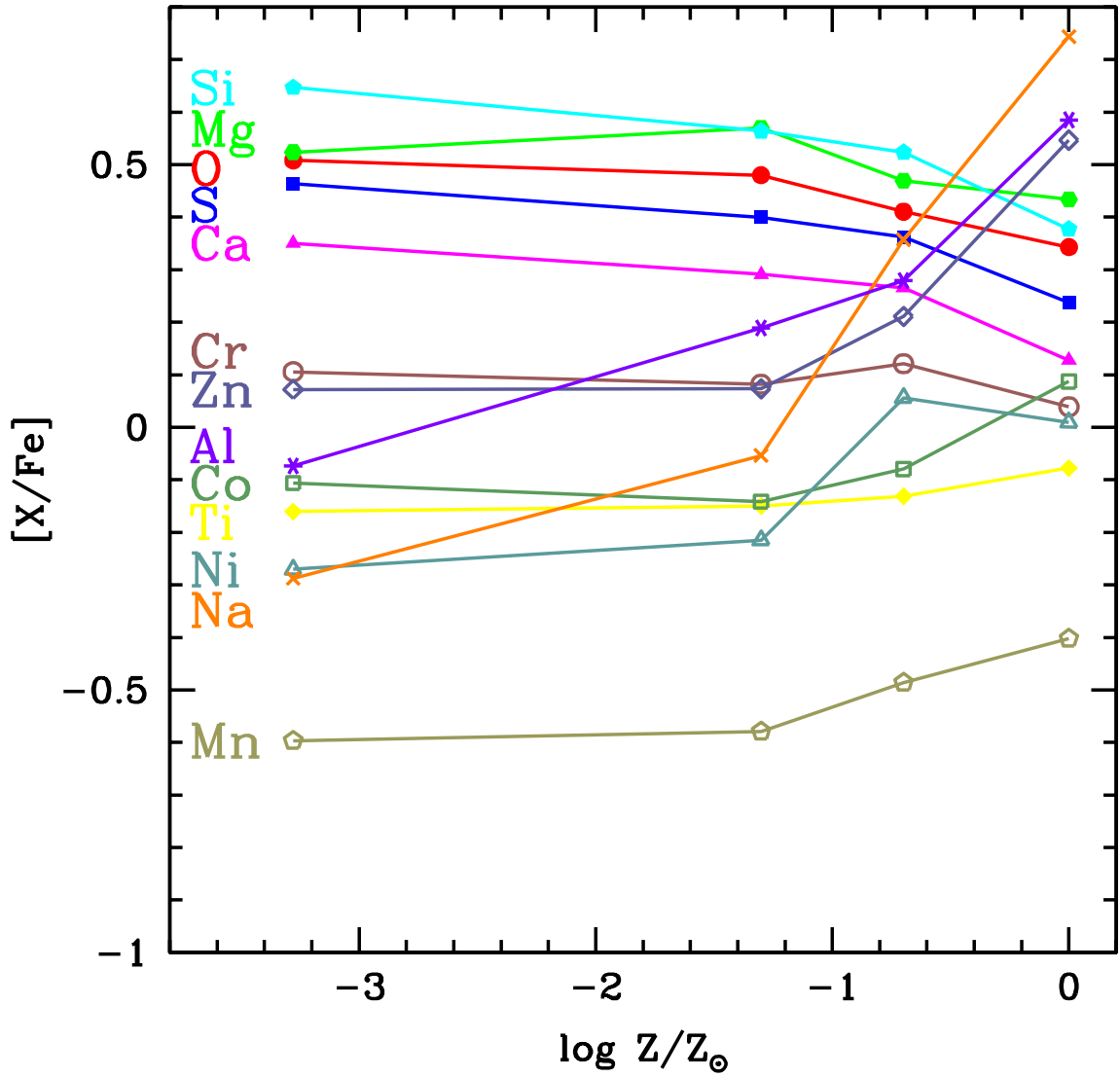}
\caption{\label{fig:yield-z}
The IMF weighted abundance ratios as a function of metallicity of progenitors, where the HN fraction $\epsilon_{\rm HN}=0.5$ is adopted.
$Z=0$ results are plotted at $\log Z=-5$.
}
\end{figure*}

Table 3 gives the IMF weighted yields as a function of metallicity, and also the SN Ia yields (W7 model from Nomoto, Thielemann, \& Yokoi 1984; Nomoto et al. 1997b) for comparison.
We adopt Salpeter IMF, i.e., a power-law mass spectrum with a slope of $x=1.35$ and a mass range from $M_\ell=0.07 M_\odot$ to $M_{\rm u}=50 M_\odot$.
We should note that the abundance ratios depend on $x$, and more strongly on $M_{\rm u}$.
Figure \ref{fig:yield-z} shows the metallicity dependence of the SN II+HN yields.
In the metal-free stellar evolution, because of the lack of initial CNO elements, the CNO cycle dose not operate until the star contracts to a much higher central temperature ($\sim 10^8$ K) than population II stars, where the 3$\alpha$ reaction produces a tiny fraction of $^{12}$C ($\sim 10^{-10}$ in mass fraction).
However, the late core evolution and the resulting Fe core masses of metal-free stars are not much different from metal-rich stars.
Therefore, the [$\alpha$/Fe] ratio is larger by only a fraction of $\sim 0.2$ dex and the abundance ratios of the iron-peak elements are not so different from metal-rich stars, except for Mn.
On the other hand, the CNO cycle produces only a small amount of $^{14}$N, which is transformed into $^{22}$Ne during He-burning. The surplus of neutrons in $^{22}$Ne increases the abundances of odd-Z elements (Na, Al, P, ...).
Therefore, the metallicity effect is realized for odd-Z elements and the inverse ratio of $\alpha$-elements and their isotopes (e.g., $^{13}$C$/^{12}$C).
[Na/Fe] and [Al/Fe] of metal-free stars are smaller by $\sim 1.0$ and $0.7$ dex than solar abundance stars, which are consistent with the observed trends (\S \ref{sec:xfe}).

Among $\alpha$-elements, it has been reported that Ca is underabundant relative to the other $\alpha$-elements in elliptical galaxies \citep{tho03}.
Although dependencies of [Ca/Fe] on the mass, metallicity, and energy are not clearly seen in our yields, [Ca/Fe] tends to be smaller for more massive and more metal-rich supernovae.
In the case of a flat IMF with rapid chemical enrichment, i.e., in ellipticals galaxies, [Ca/O] could be small.

\section{Chemical Evolution of the Solar Neighborhood}

\subsection{Chemical Evolution Model}
\label{sec:chem}

Using the one-zone chemical evolution model, we compare our nucleosynthesis yields with the observed elemental abundance ratios in the solar neighborhood.
Here the instantaneous recycling approximation is not applied, i.e., the mass dependence of yields (i.e., the stellar lifetime dependence) is taken into account, and the contributions of HNe, SNe II and SNe Ia are included, while those of low and intermediate mass stars are not.
For the solar neighborhood, we use a model that allows the infall of primordial gas from 
outside the disk region (see Kobayashi, Tsujimoto \& Nomoto 2000, hereafter K00, for the formulation).
For the infall rate, we adopt a formula
that is proportional to $t\exp[-\frac{t}{\tau_{\rm i}}]$ \citep{pag89,yos96} with an infall timescale of
$\tau_{\rm i}=5$ Gyr. 
The Galactic age is assumed to be 13 Gyr, which corresponds to the formation redshift $z_{\rm f}\sim9$ for the WMAP cosmology ($h=0.7$, $\Omega_0=0.3$, $\lambda_0=0.7$).
In K98, we assumed 15 Gyr, and the following parameters are slightly updated to meet the metallicity distribution function (MDF).
The star formation rate (SFR) is assumed to be
proportional to the gas fraction as $\psi \equiv \frac{1}{\tau_{\rm s}} f_{\rm g}$ with a timescale of $2.2$ Gyr, which is constrained from the present gas fraction $f_{\rm g}=0.15$.
The metallicity dependent main-sequence lifetime is taken from \citet{kod97}.

The treatment of SNe Ia is the same as in K98 and K00. The lifetime distribution is given by the mass ranges of companion stars in WD binary systems, which are constrained from the binary evolution \citep{hac96,hac99}.
The fraction of primary stars that eventually produce SNe Ia is given by the parameters ($b_{\rm RG}$ and $b_{\rm MS}$, respectively, for the main-sequence (MS) and red-giant (RG) companions), that are constrained from the [O/Fe]-[Fe/H] relation.
The smaller values [$b_{\rm RG}=0.02, b_{\rm MS}=0.04$] than K98 and K00 are adopted for the WD+RG and the WD+MS systems, respectively, because more Fe is produced by HNe.
The metallicity effect of SNe Ia (K98) is also taken into account. To produce SNe Ia, optically thick winds should be brown from the accreting WDs in the binary systems, which requires a large enough Fe opacity. If [Fe/H] $<-1.1$, no SN Ia can occur from these binary systems.

\begin{figure*}
\center
\includegraphics[width=16.5cm]{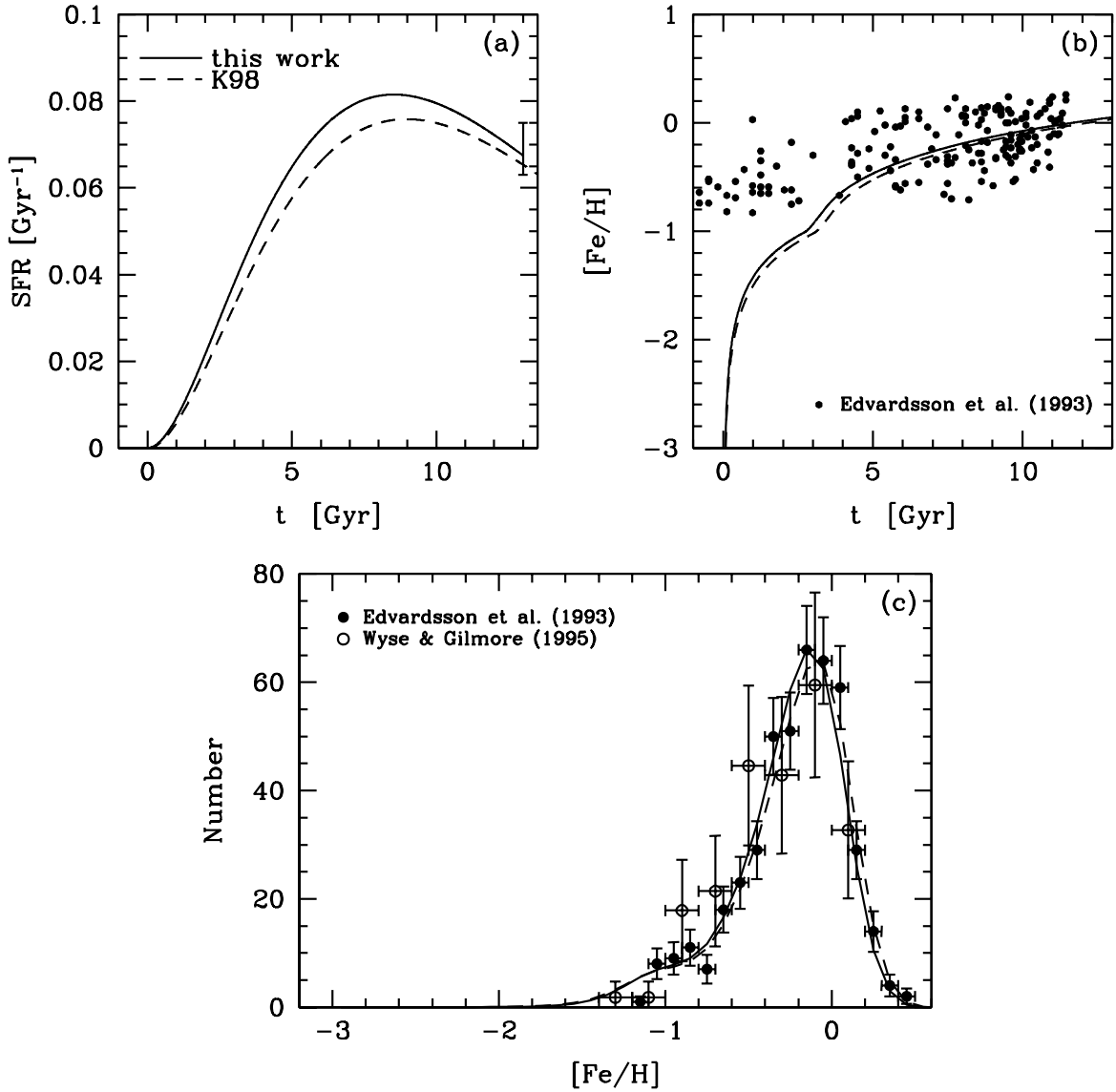}
\caption{\label{fig:mdf}
The chemical evolution of the solar neighborhood; (a) the star formation rate, (b) the age-metallicity relation, and (c) the metallicity distribution function for the model with our yields (solid line) and K98 model with N97 yields (dashed line).
Observational data sources are:
an errorbar, \citet{mat97} in panel (a);
filled circles, \citet{edv93} in panels (b) and (c); open circles, \citet{wys95} in panel (c).
}
\end{figure*}

These parameters are constrained from the chemical evolution of the solar neighborhood, and are determined as follows.
Figure \ref{fig:mdf} shows (a) the SFR, (b) the age-metallicity relation, and (c) the MDF in the solar neighborhood.
The solid and dashed lines show the results of this work and K98 model with N97 yields, respectively.
Our SFRs peak at $\sim 8$ Gyr (panel a), which is later than previous models \citep{chi97,gib03}.  The sources of this difference may be as follows:
i) We adopt the time-dependent infall rate, while others adopted the double-infall rate.
ii) The adopted parameters such as the star formation timescale are different, because these parameters are chosen in order to reproduce the present gas fraction and the MDF for different set of the adopted SN Ia model, nucleosynthesis yields, IMF, and the Galactic age.
As a result, these models can give an identical MDF, and the resultant [X/Fe]-[Fe/H] relations do not differ at all.

The iron abundance (panel b) increases quickly by SNe II and HNe at $t\ltsim3$ Gyr to reach [Fe/H] $\sim-1$, and then gradually by SNe Ia.
[Fe/H] reaches 0 at $\sim 12$ Gyr in our models, which is consistent with the average of the observational data \citep{edv93,nor04}.
This means that the solar system has formed in a relatively metal-enhanced region.
At $t\ltsim3$ Gyr, our [Fe/H] looks lower than the observations, but this is because there are many stars with unreasonably old ages in the observational data owing to the difficulty of the age estimate.
The slow accretion and slow star formation are required from the lack of metal-poor stars in the MDF (panel c).
The observed MDF \citep{edv93,wys95} is reproduced with our model by adopting $0.15$ dex convolution that corresponds to an observational error.
Recently, \citet{nor04} showed a narrower MDF. We find that our model with $0.1$ dex convolution is nearly consistent with this new MDF.
The peak metallicity of the MDF strongly depends on the slope $x$ of the IMF.
Under the above assumptions of $x$ and $M_{\rm u}$, the combination of $M_\ell$ and $b_{\rm RG}$ is constrained from the high metallicity edge of the MDF.
We choose $b$ parameters to give better agreement with the [O/Fe]-[Fe/H] evolutionary trend at [Fe/H] $\gtsim-1$.

\subsection{Evolution of [X/Fe] against [Fe/H]}
\label{sec:xfe}

\begin{figure}
\vspace*{-2.5cm}
\plotone{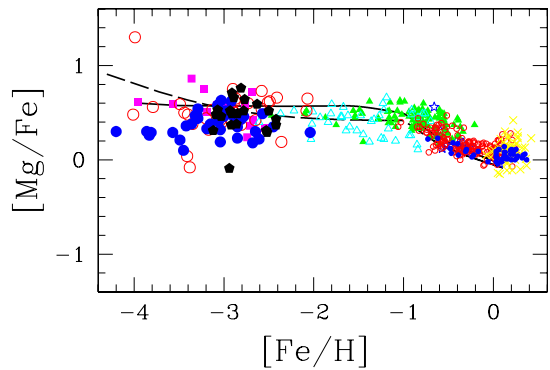}
\caption{\label{fig:mg}
[Mg/Fe]-[Fe/H] relation.
Observational data sources are:
For disk stars, small open circles, \citet{edv93};  crosses, \citet{fel98}; small filled circles, thin disk stars in \citet{ben03}; stars, filled triangles, and open triangles respectively for thin disk, dissipative component, and accretion component in \citet{gra03}.
For halo stars, large open circles, \citet{mcw95}; filled squares, \citet{rya96}; large filled circles, \citet{cay04}; filled pentagons, \citet{hon04}.
}
\end{figure}

\begin{figure}
\plotone{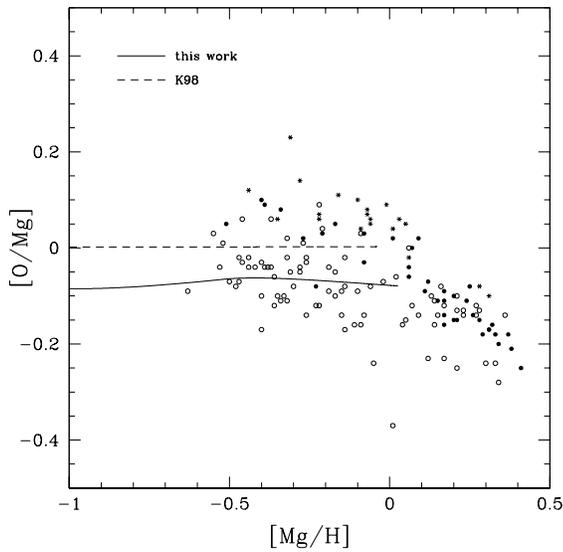}
\caption{\label{fig:mgo}
[O/Mg] against [Mg/H] for the model with our yields (solid line) and K98 model with N97 yields (dashed line).
The observational data is shown with the open circles \citep{edv93}, and the filled circle and asterisks respectively for the thin and thick disk stars \citep{ben04}.
}
\end{figure}

\begin{figure}
\vspace*{-2.5cm}
\plotone{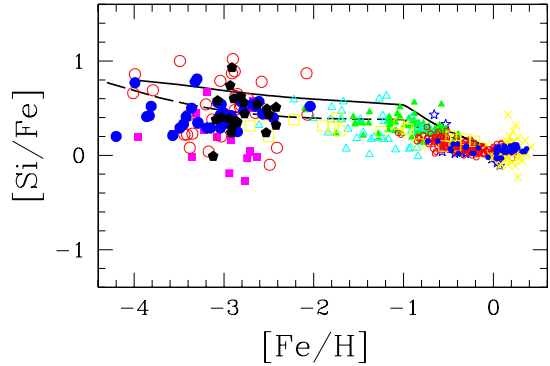}
\caption{\label{fig:si}
[Si/Fe]-[Fe/H] relation.
Observational data sources are:
For disk stars, small open circles, \citet{edv93};  crosses, \citet{fel98}; small filled circles, thin disk stars in \citet{ben03}; stars, filled triangles, and open triangles respectively for thin disk, dissipative component, and accretion component in \citet{gra03}.
For halo stars, open squares, \citet{gra91}; large open circles, \citet{mcw95}; filled squares, \citet{rya96}; large filled circles, \citet{cay04}; filled pentagons, \citet{hon04}.
}
\end{figure}

\begin{figure}
\vspace*{-2.5cm}
\plotone{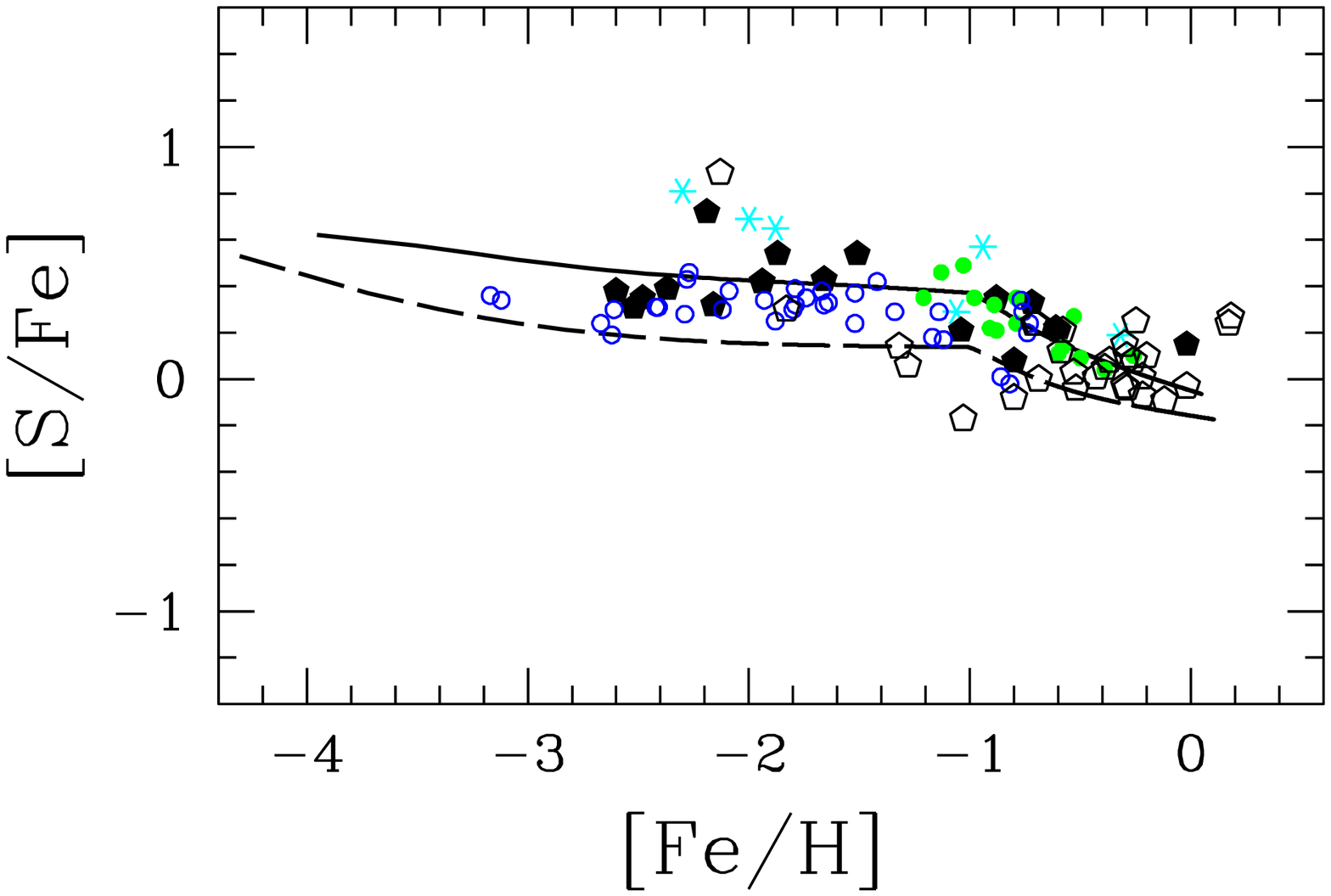}
\caption{\label{fig:s}
[S/Fe]-[Fe/H] relation.
Observational data sources are:
asterisks, \citet{isr01S}; open pentagons, \citet{tak02}; filled circles, \citet{che02}; open circles, \citet{nis04}; filled pentagons, \citet{tak05}.
}
\end{figure}

\begin{figure}
\vspace*{-2.5cm}
\plotone{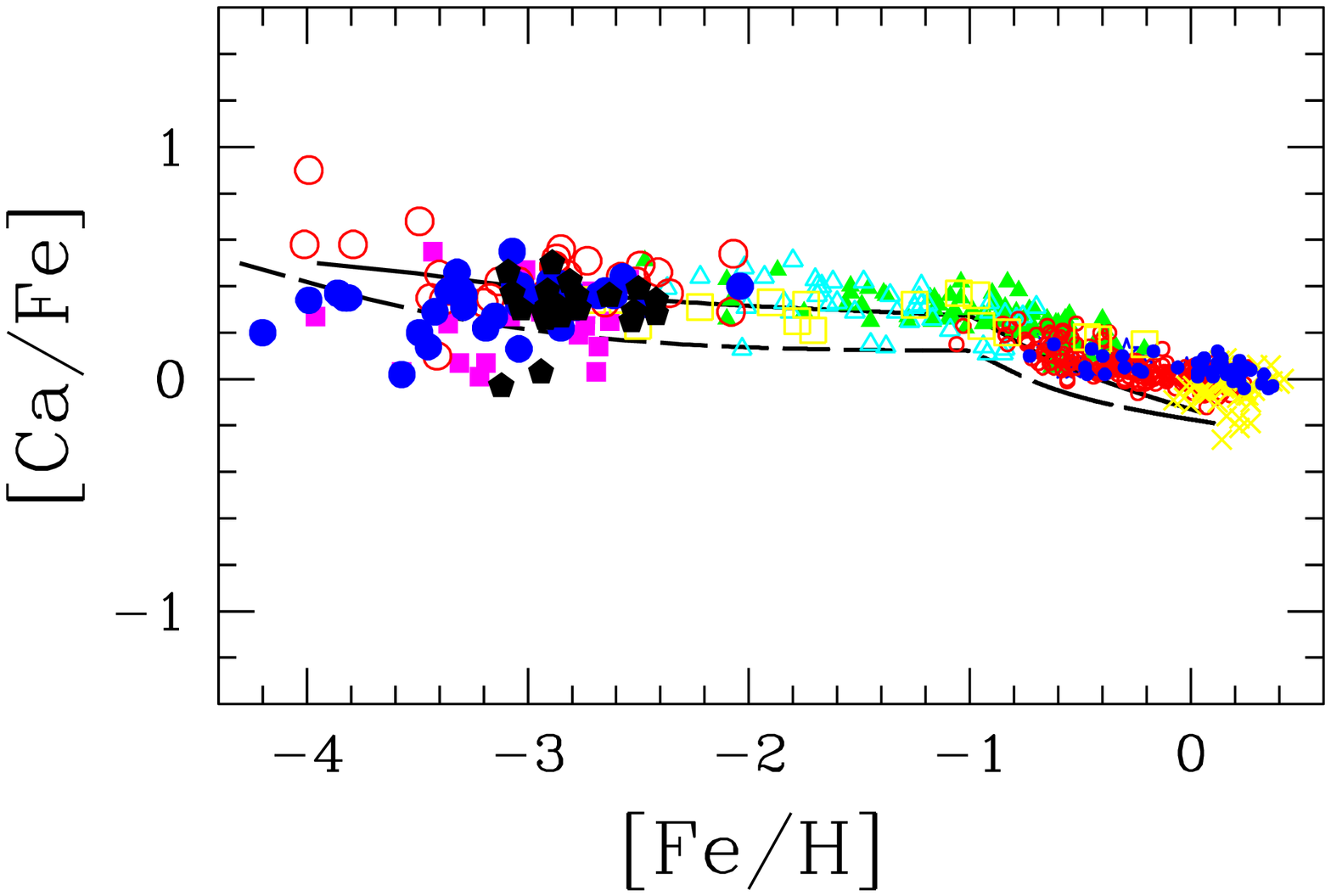}
\caption{\label{fig:ca}
[Ca/Fe]-[Fe/H] relation.
See Fig.\ref{fig:si} for the observational data sources.
}
\end{figure}

\begin{figure}
\vspace*{-2.5cm}
\plotone{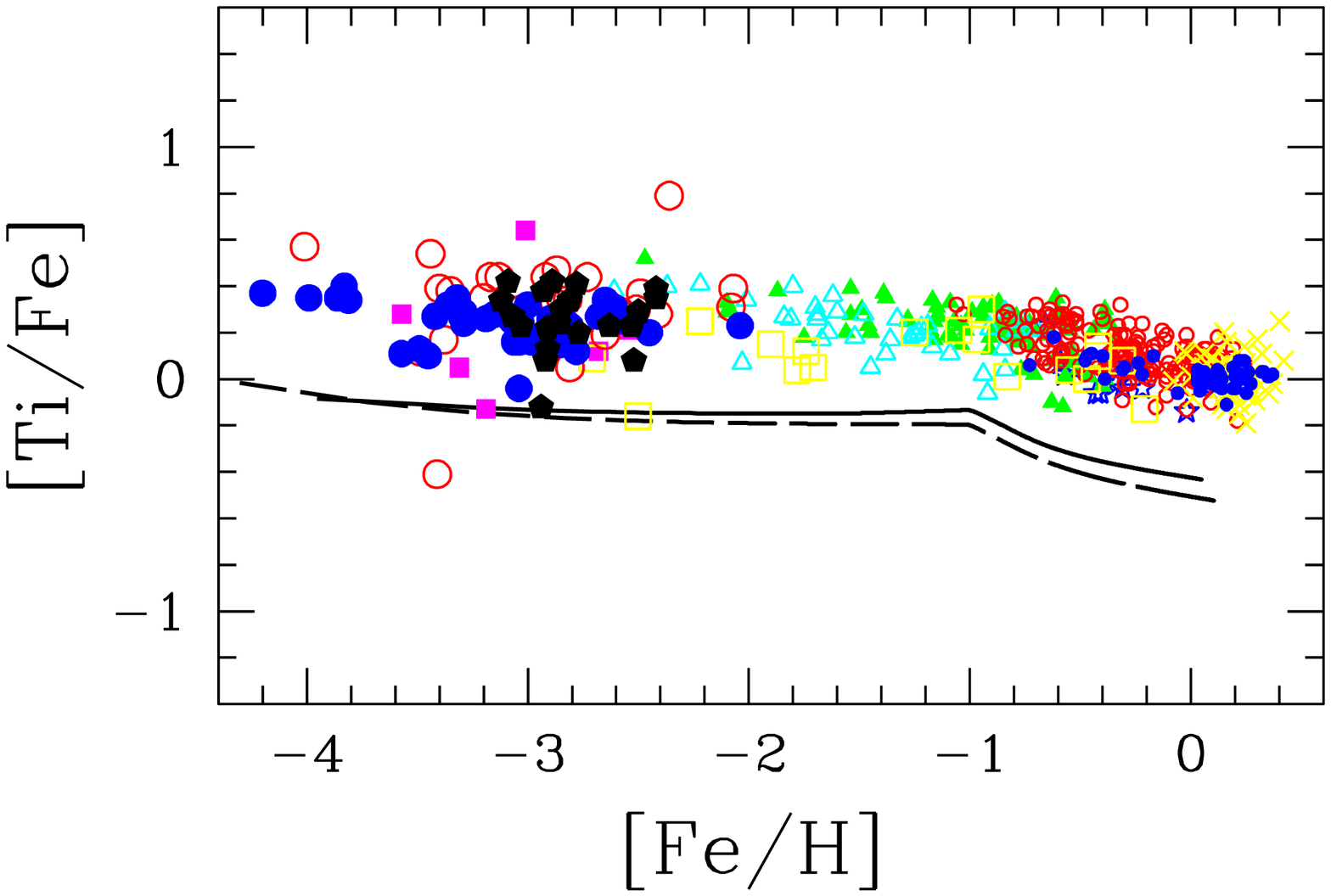}
\caption{\label{fig:ti}
[Ti/Fe]-[Fe/H] relation.
See Fig.\ref{fig:si} for the observational data sources.
}
\end{figure}

\begin{figure}
\vspace*{-2.5cm}
\plotone{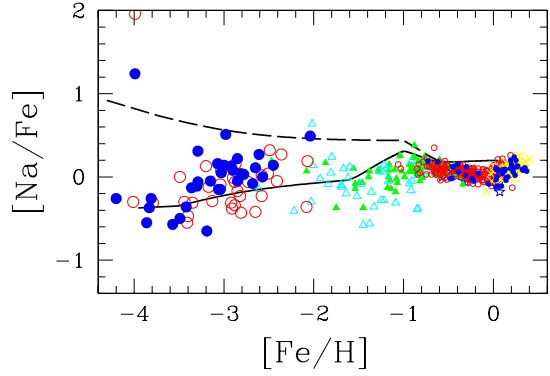}
\caption{\label{fig:na}
[Na/Fe]-[Fe/H] relation.
Observational data sources are:
For disk stars, small open circles, \citet{edv93}; crosses, \citet{fel98}; small filled circles, thin disk stars in \citet{ben03}; stars, filled triangles, and open triangles respectively for thin disk, dissipative component, and accretion component in \citet{gra03}.
For halo stars, large open circles, \citet{mcw95}; large filled circles, \citet{cay04}.
}
\end{figure}

\begin{figure}
\vspace*{-2.5cm}
\plotone{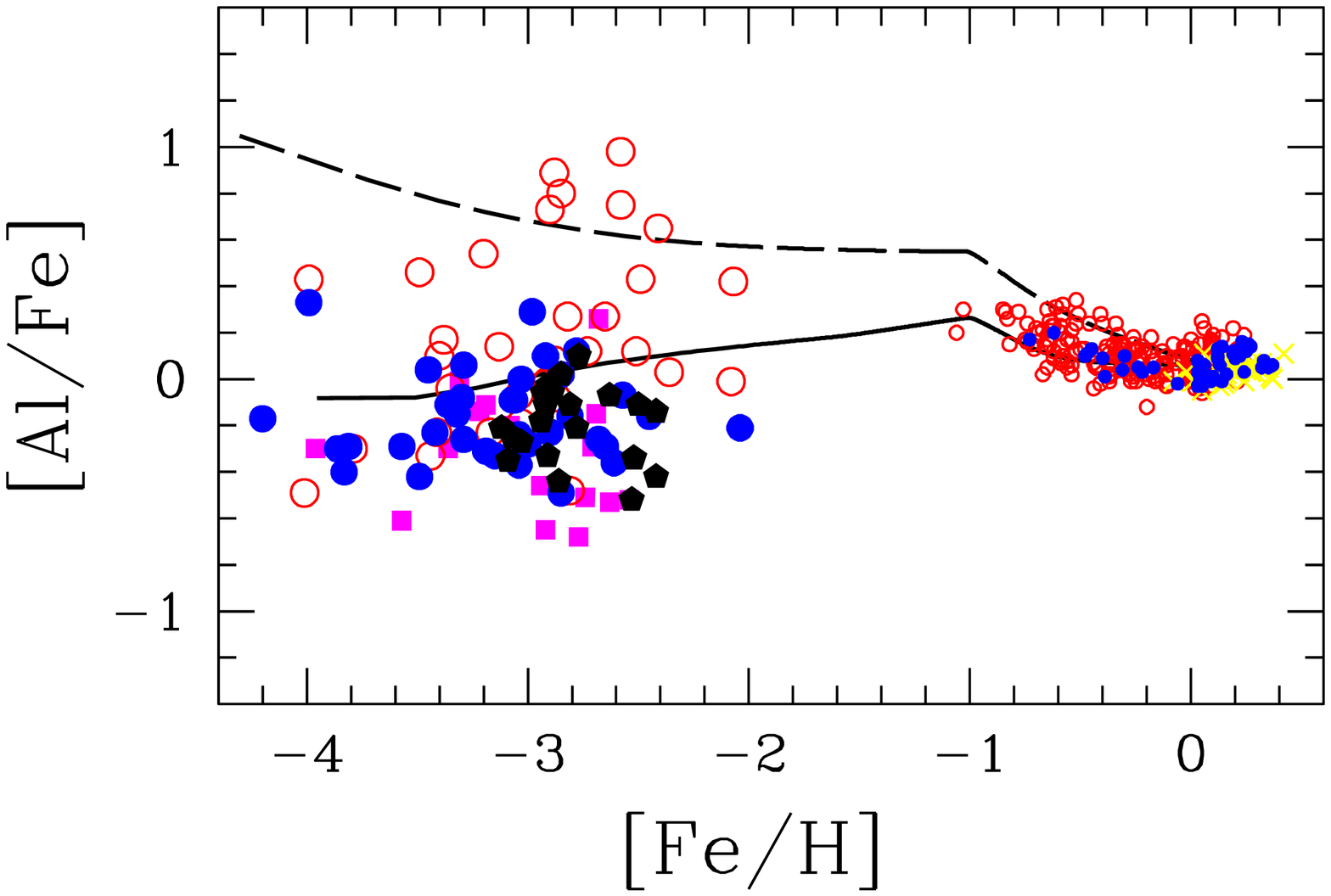}
\caption{\label{fig:al}
[Al/Fe]-[Fe/H] relation.
Observational data sources are:
For disk stars, small open circles, \citet{edv93}; crosses, \citet{fel98}; small filled circles, thin disk stars in \citet{ben03}.
For halo stars, large open circles, \citet{mcw95}; filled squares, \citet{rya96}; large filled circles, \citet{cay04}; filled pentagons, \citet{hon04}.
}
\end{figure}

\begin{figure}
\vspace*{-2.5cm}
\plotone{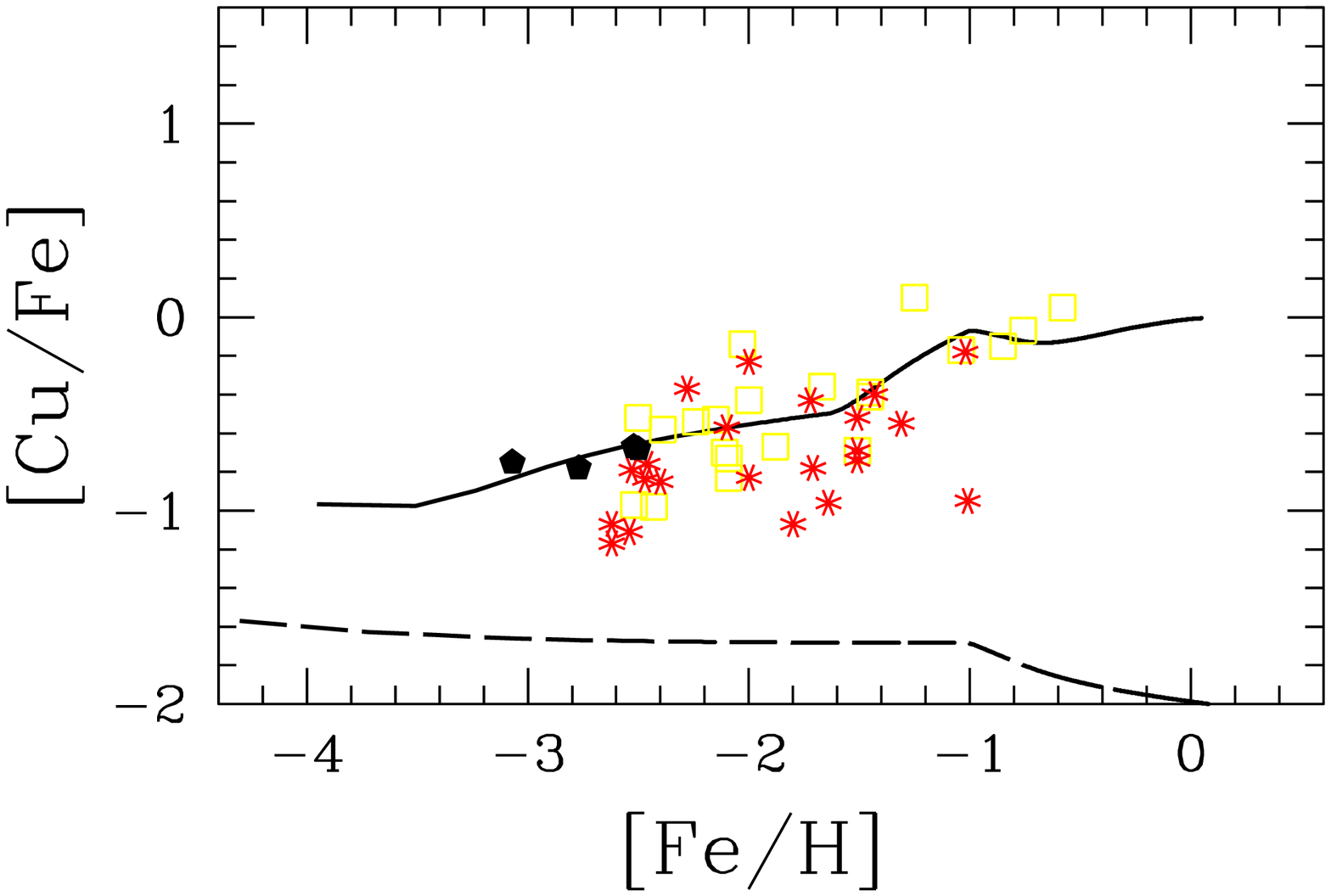}
\caption{\label{fig:cu}
[Cu/Fe]-[Fe/H] relation.
Observational data sources are:
For halo stars, open squares, \citet{sne91}; eight-pointed asterisks, \citet{pri00}; filled pentagons, \citet{hon04}.
}
\end{figure}

\begin{figure}
\vspace*{-2.5cm}
\plotone{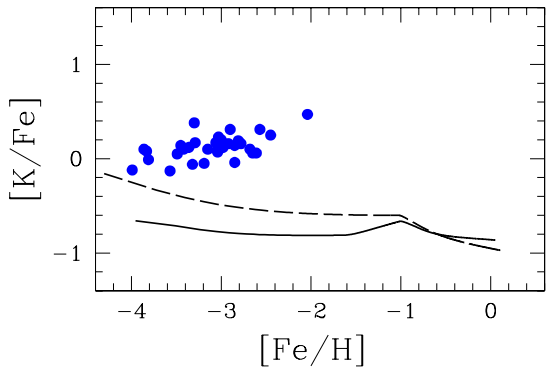}
\caption{\label{fig:k}
[K/Fe]-[Fe/H] relation.
Observational data source is:
For halo stars, large filled circles, \citet{cay04}.
}
\end{figure}

\begin{figure}
\vspace*{-2.5cm}
\plotone{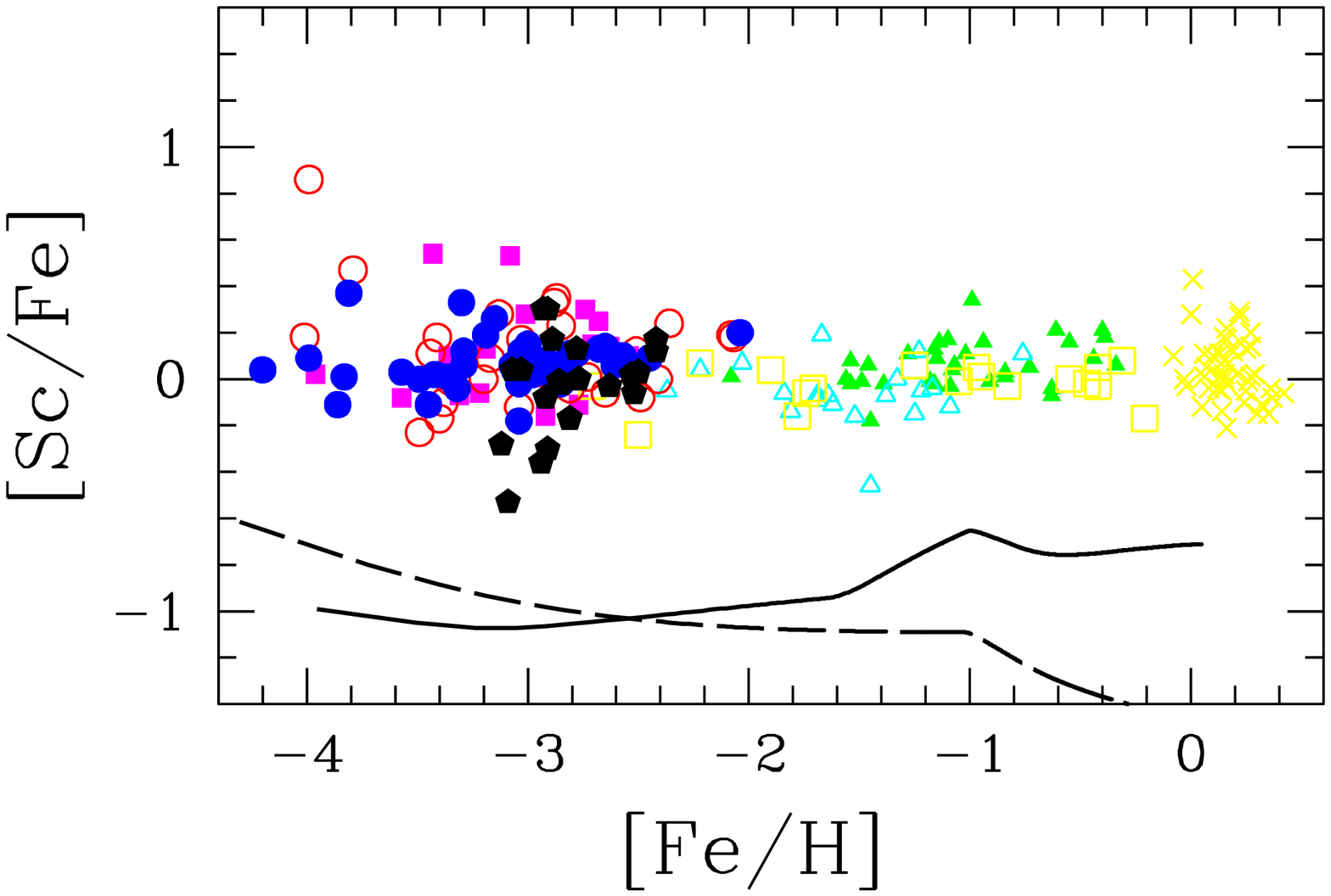}
\caption{\label{fig:sc}
[Sc/Fe]-[Fe/H] relation.
Observational data sources are:
For disk stars, crosses, \citet{fel98}; small filled circles, thin disk stars in \citet{ben03}; filled and open triangles respectively for dissipative component and accretion component in \citet{gra03}.
For halo stars, open squares, \citet{gra91}; large open circles, \citet{mcw95}; filled squares, \citet{rya96}; large filled circles, \citet{cay04}; filled pentagons, \citet{hon04}.
}
\end{figure}

\begin{figure}
\vspace*{-2.5cm}
\plotone{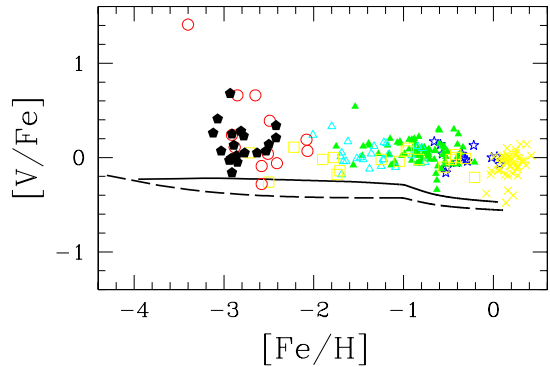}
\caption{\label{fig:v}
[V/Fe]-[Fe/H] relation.
Observational data sources are:
For disk stars, crosses, \citet{fel98}; stars, filled triangles, and open triangles respectively for thin disk, dissipative component, and accretion component in \citet{gra03}.
For halo stars, open squares, \citet{gra91}; large open circles, \citet{mcw95}; filled pentagons, \citet{hon04}.
}
\end{figure}

\begin{figure}
\vspace*{-2.5cm}
\plotone{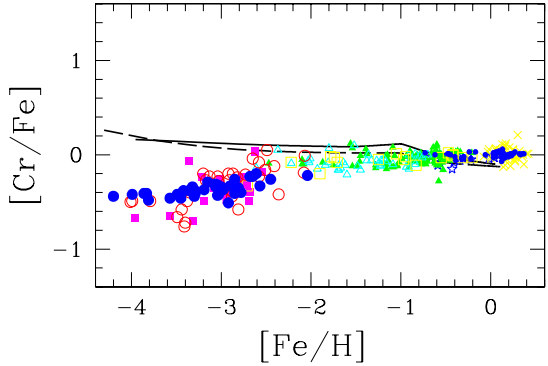}
\caption{\label{fig:cr}
[Cr/Fe]-[Fe/H] relation.
Observational data sources are:
For disk stars, crosses, \citet{fel98}; small filled circles, thin disk stars in \citet{ben03}; stars, filled triangles, and open triangles respectively for thin disk, dissipative component, and accretion component in \citet{gra03}.
For halo stars, open squares, \citet{gra91}; large open circles, \citet{mcw95}; filled squares, \citet{rya96}; large filled circles, \citet{cay04}.
}
\end{figure}

\begin{figure}
\vspace*{-2.5cm}
\plotone{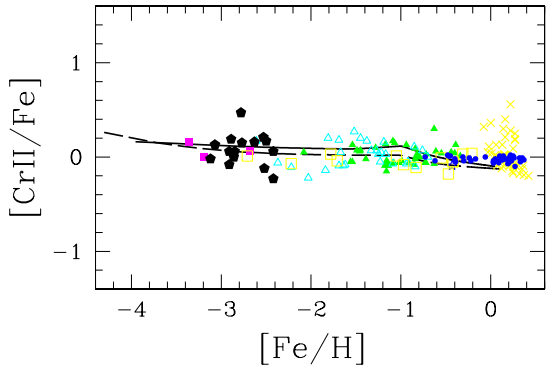}
\caption{\label{fig:cr2}
[CrII/Fe]-[Fe/H] relation.
Observational data sources are:
For disk stars, crosses, \citet{fel98}; small filled circles, thin disk stars in \citet{ben03}; stars, filled triangles, and open triangles respectively for thin disk, dissipative component, and accretion component in \citet{gra03}.
For halo stars, open squares, \citet{gra91}; filled squares, \citet{rya96}; filled pentagons, \citet{hon04}.
}
\end{figure}

\begin{figure}
\vspace*{-2.5cm}
\plotone{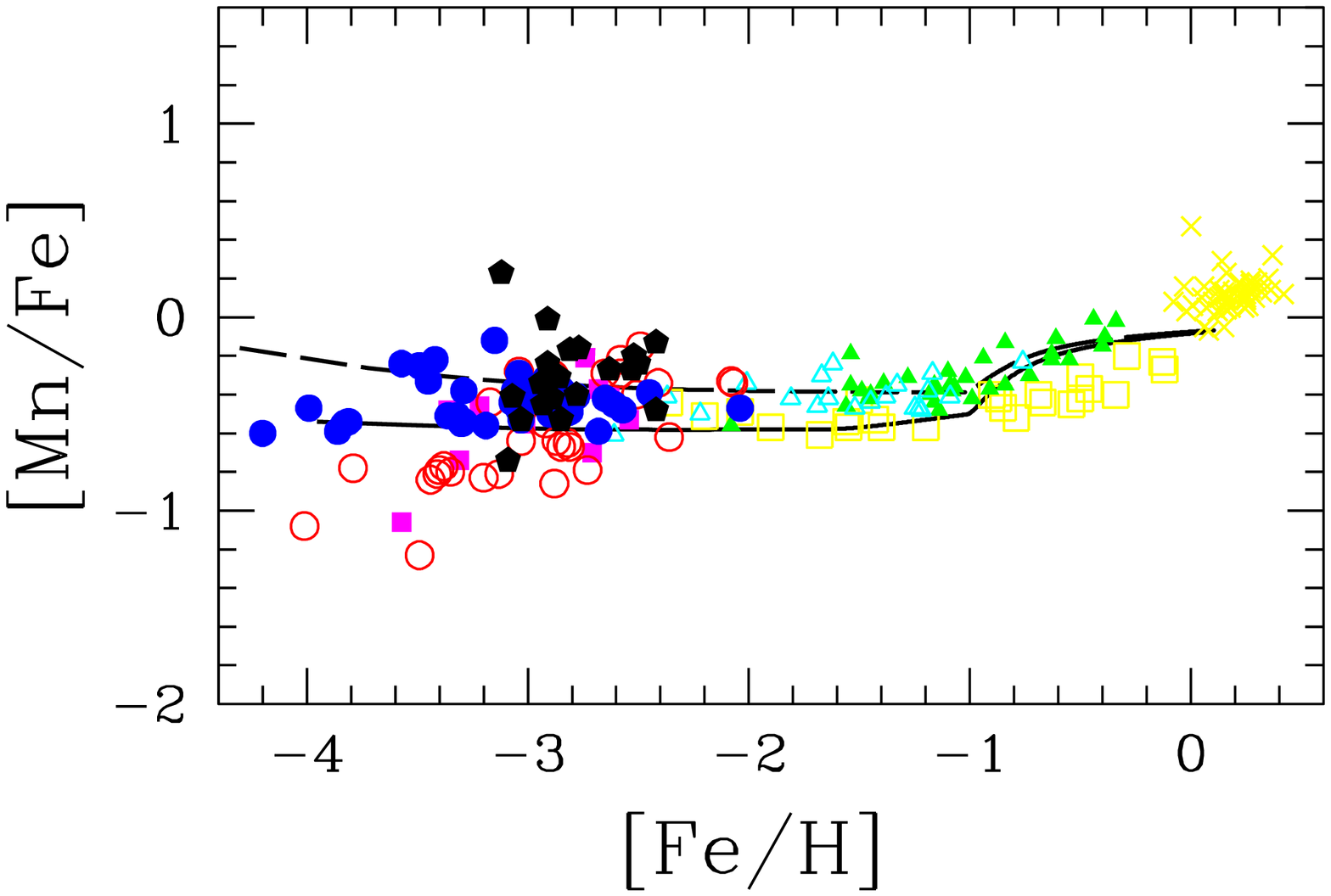}
\caption{\label{fig:mn}
[Mn/Fe]-[Fe/H] relation.
Observational data sources are:
For disk stars, crosses, \citet{fel98}; filled and open triangles respectively for dissipative component and accretion component in \citet{gra03}.
For halo stars, open squares, \citet{gra89}; large open circles, \citet{mcw95}; filled squares, \citet{rya96}; large filled circles, \citet{cay04}; filled pentagons, \citet{hon04}.
}
\end{figure}

\begin{figure}
\vspace*{-2.5cm}
\plotone{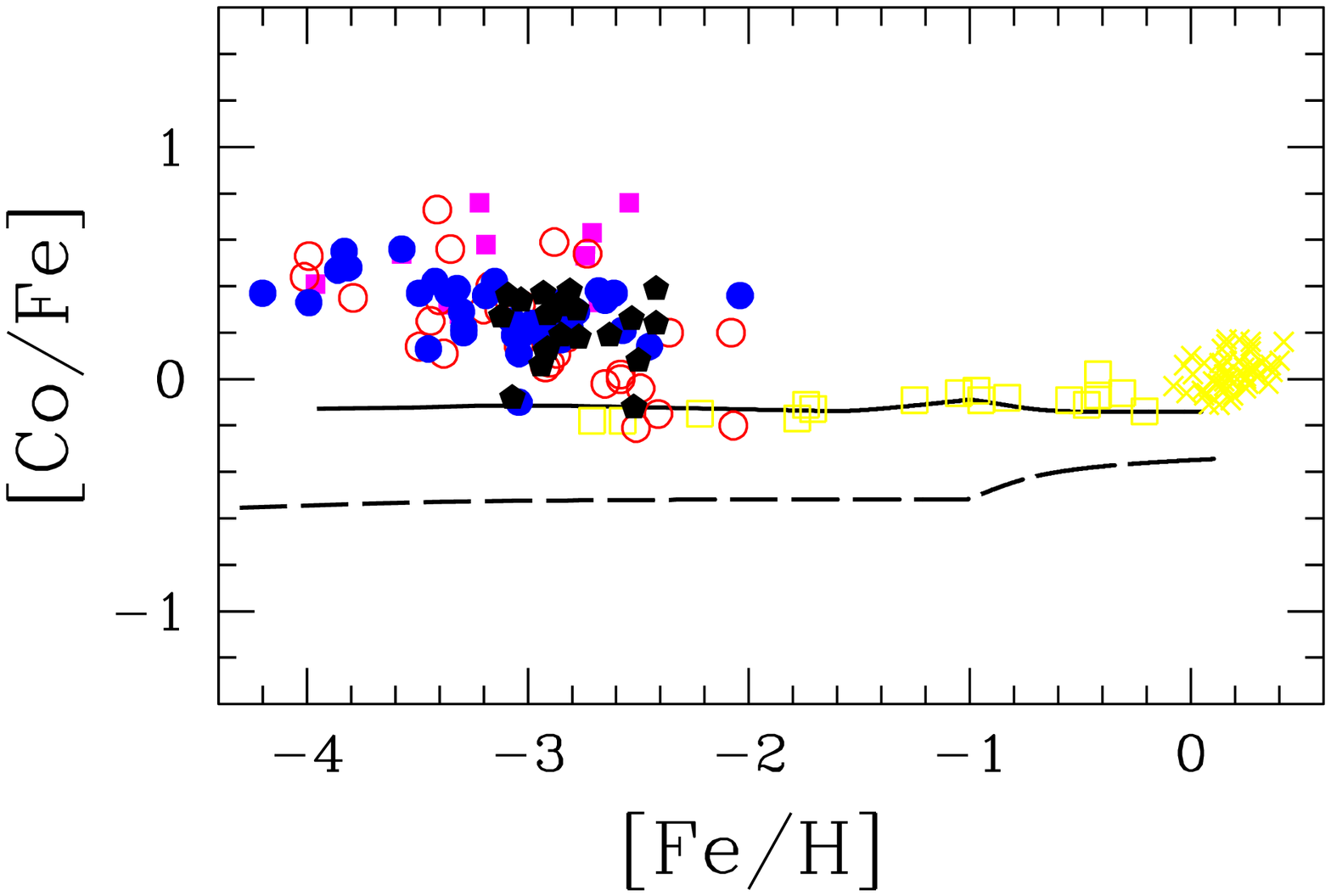}
\caption{\label{fig:co}
[Co/Fe]-[Fe/H] relation.
Observational data sources are:
For disk stars, crosses, \citet{fel98}.
For halo stars, open squares, \citet{gra91}; large open circles, \citet{mcw95}; filled squares, \citet{rya96}; large filled circles, \citet{cay04}; filled pentagons, \citet{hon04}.
}
\end{figure}

\begin{figure}
\vspace*{-2.5cm}
\plotone{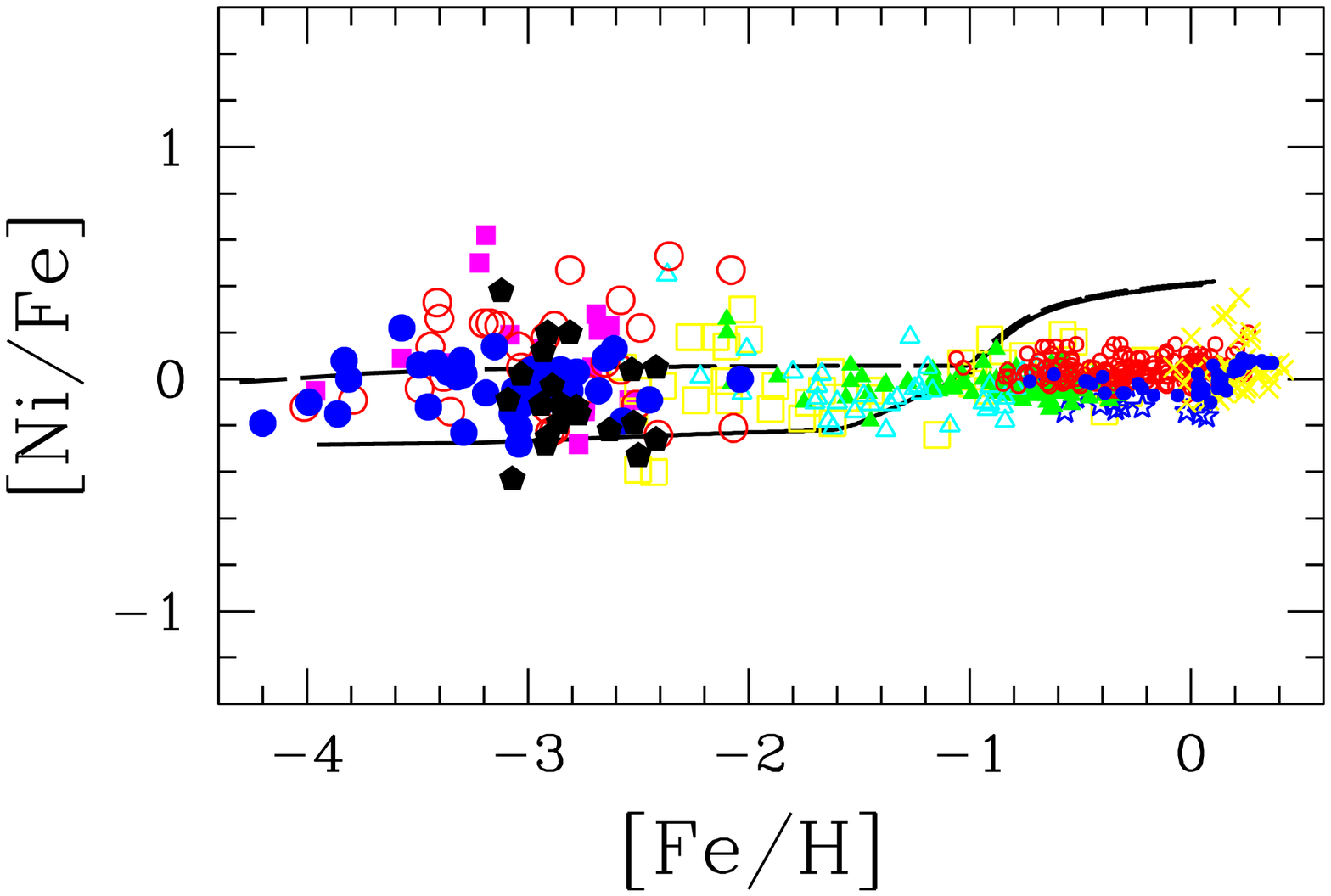}
\caption{\label{fig:ni}
[Ni/Fe]-[Fe/H] relation.
Observational data sources are:
For disk stars, small open circles, \citet{edv93}; crosses, \citet{fel98}; small filled circles, thin disk stars in \citet{ben03}; stars, filled triangles, and open triangles respectively for thin disk, dissipative component, and accretion component in \citet{gra03}.
For halo stars, open squares, \citet{sne91}; large open circles, \citet{mcw95}; filled squares, \citet{rya96}; large filled circles, \citet{cay04}; filled pentagons, \citet{hon04}.
}
\end{figure}

\begin{figure}
\vspace*{-2.5cm}
\plotone{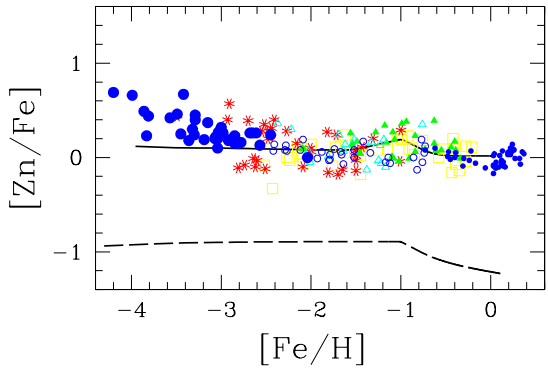}
\caption{\label{fig:zn}
[Zn/Fe]-[Fe/H] relation.
Observational data sources are:
For disk stars, small filled circles, thin disk stars in \citet{ben03}; filled and open triangles respectively for dissipative component, and accretion component in \citet{gra03}.
For halo stars, open squares, \citet{sne91}; eight-pointed asterisks, \citet{pri00}; open circles, \citet{nis04}; large filled circles, \citet{cay04}.
}
\end{figure}

\begin{figure}
\vspace*{-2.5cm}
\plotone{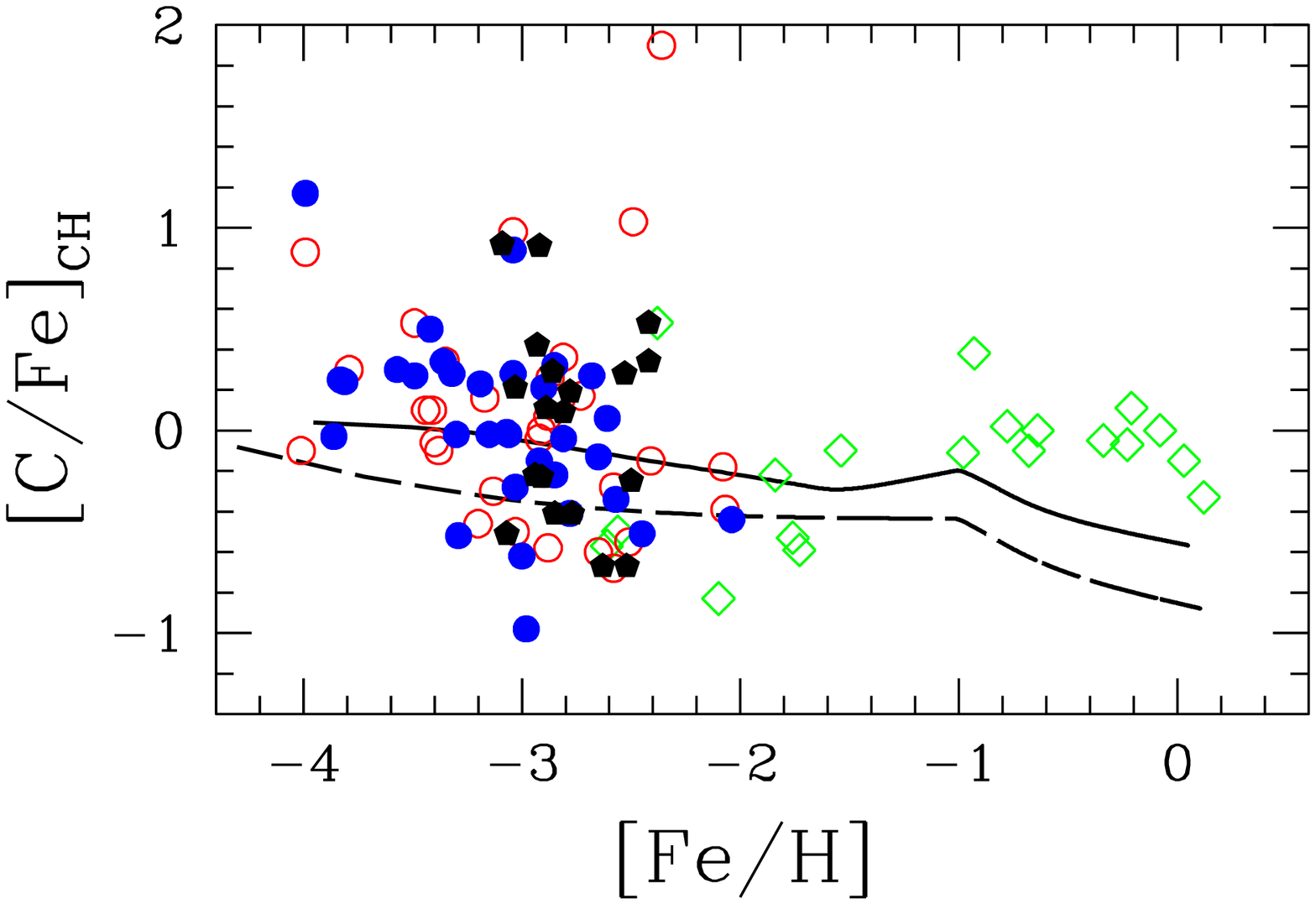}
\caption{\label{fig:c}
[C/Fe]-[Fe/H] relation.
Observational data sources are:
For disk stars, open diamonds, \citet{car00}.
For halo stars, large open circles, \citet{mcw95}; large filled circles, \citet{cay04}; filled pentagons, \citet{hon04}.
}
\end{figure}

\begin{figure}
\vspace*{-2.5cm}
\plotone{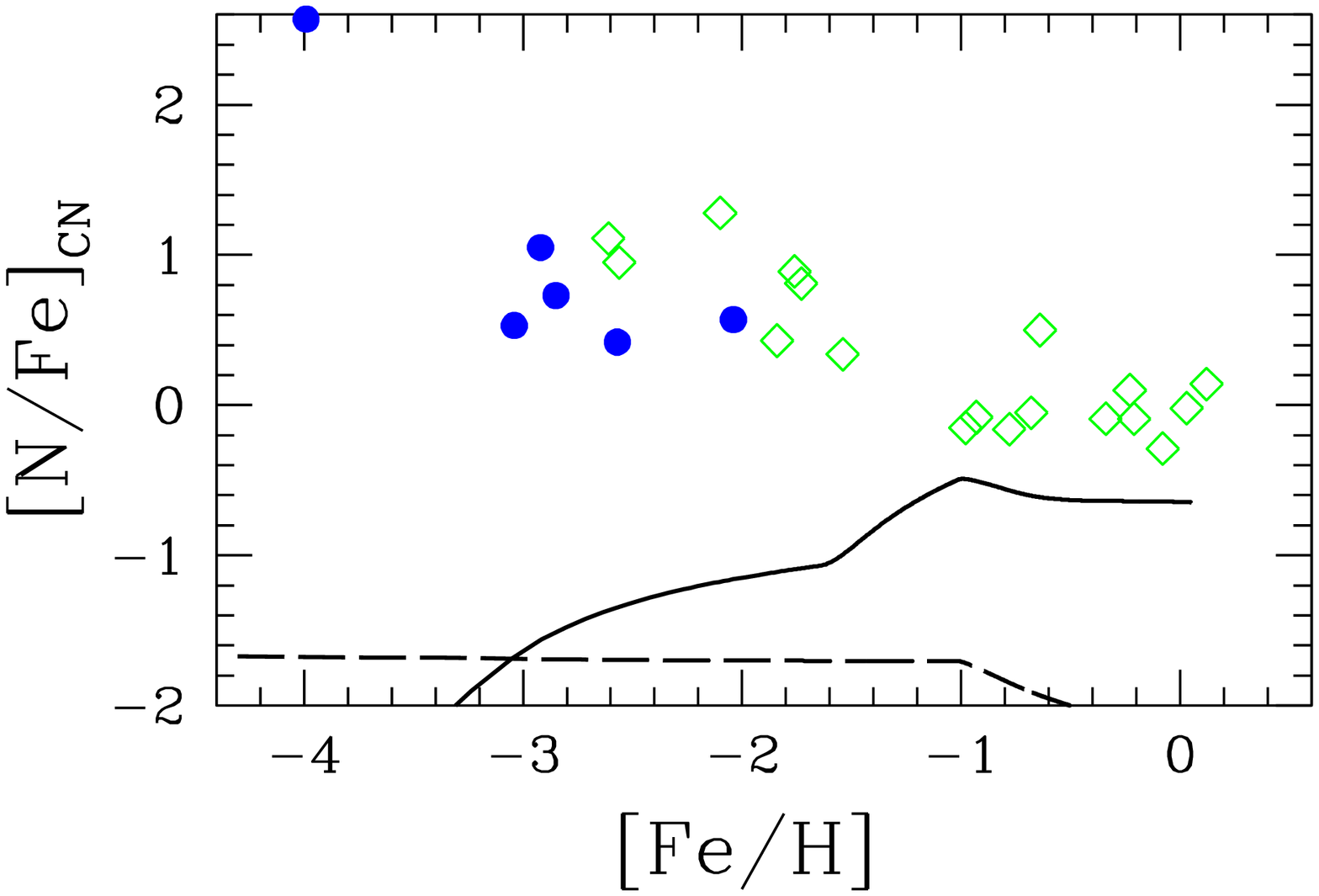}
\caption{\label{fig:n}
[N/Fe]-[Fe/H] relation.
Observational data sources are:
For disk stars, open diamonds, \citet{car00}.
For halo stars, large filled circles, \citet{cay04}.
}
\end{figure}

Figures \ref{fig:o}-\ref{fig:xfe} show the evolutions of heavy element abundance ratios [X/Fe] against [Fe/H]. 
Our model with new yields (solid line) are in much better agreement with the observational data than K98 model with N97 yields (dashed line), especially for Al, Na, Ca, and Zn.
As the time goes, the iron abundance increases, and the abundance ratio for many elements stays constant with a plateau value at [Fe/H] $\ltsim-1$, which is determined only by SNe II and HNe.
From [Fe/H] $\sim-1$, SNe Ia start to occur (see K98 for SN Ia models) producing more Fe than $\alpha$-elements, and thus [$\alpha$/Fe] decreases toward the solar abundance.
In the following subsections, we discuss the details for each element.

\subsubsection{Oxygen}

O is the most abundant heavy element that covers a half of metallicity for the solar abundance and is one of the best described elements in nucleosynthesis.
However, the observation has been debated. 
i) The abundances determined from the forbidden line [OI] at $6300 \AA$ in giants show a plateau with [O/Fe] $\sim 0.4-0.5$. ii) Those from the near-IR triplet OI at $7774 \AA$ in unevolved subdwarfs suggest that [O/Fe] gradually increases with decreasing [Fe/H] to reach [O/Fe] $\sim0.8$ at [Fe/H] $\sim-3$. iii) Those from the OH line in the near-UV of unevolved stars show a monotonic increase with a steeper slope from [Fe/H] $\sim$ [O/Fe] $\sim0$ to [Fe/H] $\sim-3$ and [O/Fe] $\sim1$ (Israelian et al. 1998, 2001; Boesgaard et al. 1999).
However, when a suitable temperature scale is adopted and the NLTE and 3D effects are taken into account, [OI], OI, and IR OH lines give consistent results. A plateau of [O/Fe] $\sim0.3-0.45$ is seen at $-2\ltsim$ [Fe/H] $\ltsim-1$: [O/Fe] $\sim 0.45$ \citep{car00}, 0.35 \citep{mel02}, and 0.3 with 3D correction \citep{nis02}.
At [Fe/H] $\sim-3$, a gentle increase in [O/Fe] $\sim0.5-0.6$ is seen: 0.47 with 3D correction \citep{cay04}.

\begin{figure}
\vspace*{-2.5cm}
\plotone{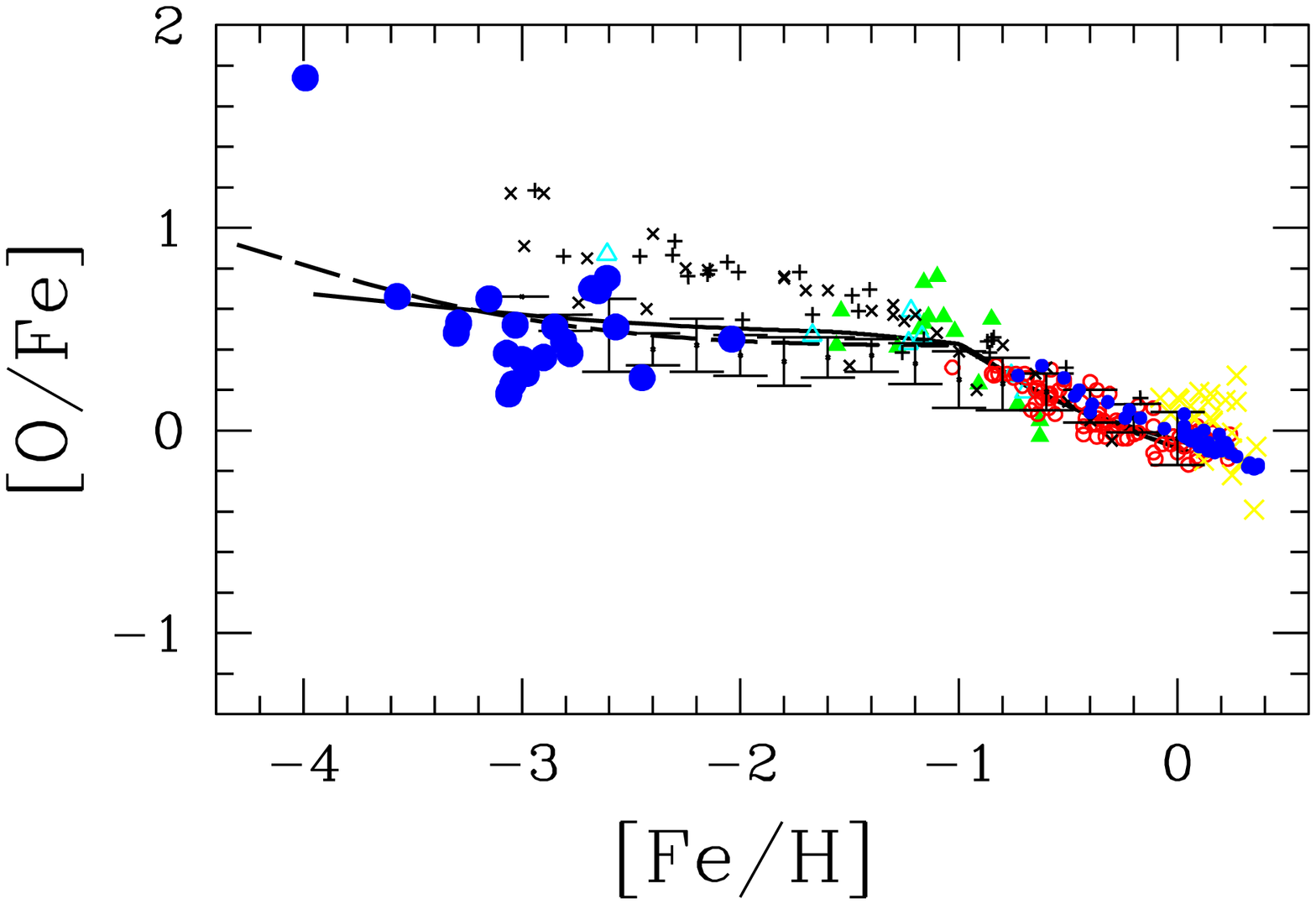}
\caption{\label{fig:o}
[O/Fe]-[Fe/H] relation for the model with our yields (solid line) and K98 model with N97 yield (dashed line).
Observational data sources are:
For disk stars, small open circles, \citet{edv93};  crosses, \citet{fel98}; small filled circles, thin disk stars in \citet{ben04}; filled and open triangles respectively for dissipative component and accretion component in \citet{gra03}.
For halo stars, large filled circles, \citet{cay04}.
UV OH observation is shown with small crosses \citep[2001]{isr98} and plus \citep{boe99}.
The errorbars shows the average taken with [OI] from \citet{mel02}.
}
\end{figure}

\begin{figure*}
\center
\includegraphics[width=16.5cm]{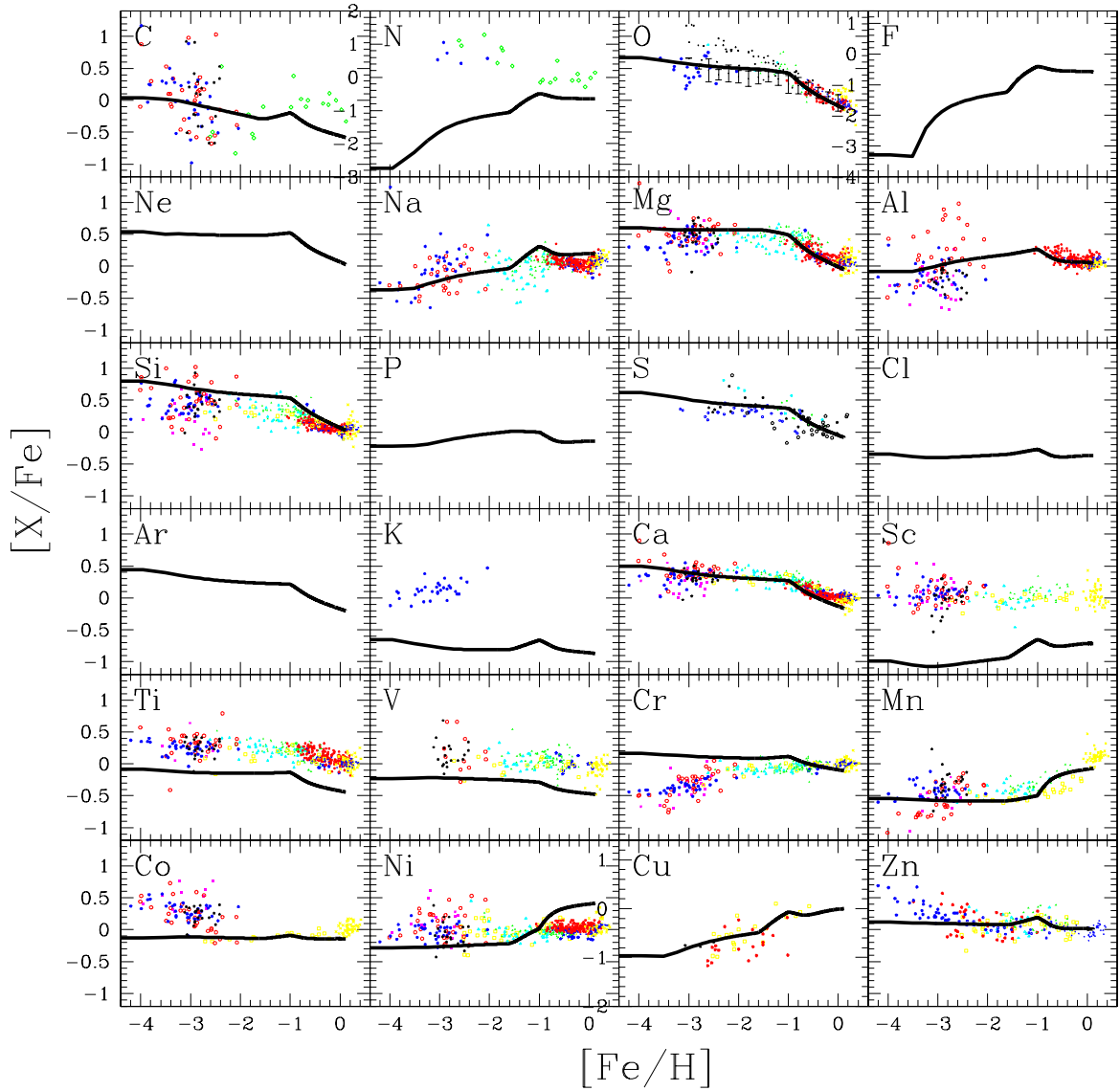}
\caption{\label{fig:xfe}
[X/Fe]-[Fe/H] relations.
See Figs.\ref{fig:o}-\ref{fig:n} for the observational data sources.
}
\end{figure*}

C, N, O, Ne, and Mg are mainly produced in hydrostatic burning phase, so that their yields depend mainly on the pre-supernova model.
In our model, [O/Fe] is $0.42$ at [Fe/H] $=-1$ and slightly increases to $0.57$ at [Fe/H] $=-3$, being consistent with the observations (except for UV OH results) as shown in Figure \ref{fig:o}.
The gradual increase in [O/Fe] with decreasing [Fe/H] stems from larger [O/Fe] in more massive SNe II, more metal-poor SNe II, and HNe.
The metallicity dependence is as small as $0.15$ dex between $Z=Z_\odot$ and $Z=0$ (Fig.\ref{fig:yield-z}). The mass dependence is large for normal SNe II ([O/Fe] $\sim -0.5$ to $1$ for $Z=Z_\odot$; see Figs.\ref{fig:yield}-\ref{fig:yield4}), but such a dependence is weakened by the HN contribution because HNe produce more Fe and give constant [O/Fe].
From [Fe/H] $\sim -1$, [O/Fe] decreases quickly due to a large amount of Fe production by SNe Ia.

Low [$\alpha$/Fe] is often used to discuss the formation timescale under the assumption that the SN Ia lifetime is 1.5 Gyr.
We note, however, this approach would be misleading if the following effects are not taken into account: the lifetime distribution and the metallicity effect of SNe Ia, the mass and energy dependences of the nucleosynthesis yields of SNe II and HNe, and an uncertainty of the IMF.
i) The enrichment by SNe Ia results in low [$\alpha$/Fe]. The shortest lifetime of SNe Ia depends on the SN Ia model: $\sim 0.1$ Gyr for the double-degenerate \citep{tut94}, $\sim 0.3$ Gyr for \citet{mat01}'s model, and $\sim 0.5$ Gyr for our single-degenerate model (K98).
ii) Some anomalous stars \citep[e.g., ][]{nis97} have [O/Fe] $\sim$ 0 at [Fe/H] $\ltsim -1$. Such small [O/Fe] can be explained by the low-mass $13-15M_\odot$ SNe II, where O yield is smaller than in massive stars (K00).
The abundance patterns in the dwarf spheroidal galaxies can be explained these SNe II \citep{tol03,tra04}.
iii) HNe may produce even a larger amount of Fe due to a smaller fallback mass or a larger energy than the typical HNe in our yields.
iv) Also SNe I.5, which are the SN Ia-like explosions of metal-poor AGB stars, have been suggested by \citet{nom03} \citep[see also][]{tsu06}.

On the other hand, very large [O/Fe] ($>1$) could be explained by the small Fe production, namely by either i) massive SNe II with $E_{51} \le 1$ or ii) HNe with a larger mass-cut (i.e., a large black-hole mass).
In such stars, other $\alpha$-elements should show the same trend, especially Mg, Si, and S.
The difference between these two possibilities appears in the abundance ratios among the iron-peak elements, especially large [(Zn,Co)/Fe].
CS 22949-037 with [O/Fe] $\sim2$ at [Fe/H] $\sim -4$ in Figure \ref{fig:o} has large [Mg/Fe] ($\sim 1.6$), normal [(Ca,Ti)/Fe] ($\sim0.35$), and large [(Zn,Co)/Fe], which are explained with a HN model (UN05). 
The low-energy explosion with $E_{51}=0.4$ in Tsujimoto \& Shigeyama (2003) also gives large [O/Fe] but cannot explain the large [(Zn,Co)/Fe].

\subsubsection{Magnesium}

Mg is one of the best observed elements with several lines and little NLTE effect.
Cayrel et al. (2004, hereafter C04) claimed that [(Mg,Si,Ca,Ti)/Fe] is constant as $\sim 0.2-0.3$ with a very small dispersion of $\sim 0.1$ dex.
In our chemical evolution model, [Mg/Fe] is $0.49$ at [Fe/H] $\sim-1$ and slightly increases to $0.57$ at [Fe/H] $\sim-3$, which is larger than $0.27$ in C04, but in good agreement with the observational data over the wide range of [Fe/H] as shown in Figure \ref{fig:mg}.
We note that the overall agreement of the [Mg/Fe]-[Fe/H] relation is due to our assumptions of the mixing-fallback, the HN fraction $\epsilon_{\rm HN}=0.5$, the upper mass limit $M_{\rm u}=50M_\odot$, and the time and metallicity independent IMF.
SNe II typically provide [Mg/Fe] $\sim0.5$ varying between $-0.2$ ($Z=Z_\odot, 18M_\odot$) and $1$ ($40M_\odot$).
For HNe, [Mg/Fe] $\sim 0.5$ despite the large progenitor masses ($M\ge20M_\odot$) because the iron yield is as large as the yields of $\alpha$-elements.
Therefore, the scatter of [Mg/Fe] can be small independent of the mixing process of interstellar medium \citep{tom06,nom06}.

[O/Mg] is $\sim0$ independent of [Fe/H] in N97 yields, while it is in a range from $\sim -0.08$ to $-0.06$ in our new yields.
Compared with Edvardsson et al. (1993)'s data for [Fe/H] $\gtsim-1$, there was 0.1 dex offset in [O/Mg] with N97 yields, of which problem is solved with our yields (Fig.\ref{fig:mgo}).
Recently, however, Bensby et al. (2004) showed 0.1 dex larger [O/Mg] for [Mg/H] $\ltsim0$ and decreasing trend for high metallicity.
With our model, the iron abundance does not reach such high metallicity, and we do not calculate the nucleosynthesis yields for $Z>Z_\odot$ yet.
Such a decrease in O/Mg would require some additional effects which are not included in our stellar evolution models such as strong stellar winds or a process that causes the change in the C/O ratio.

\subsubsection{Silicon, Sulfur, Calcium, and Titanium}

The observed Si abundance is represented by only two lines and affected by the contamination of CH and H$\delta$ lines. These may arise the larger scatter than Mg.
[Si/Fe] is $0.53-0.68$ at $-3\ltsim$ [Fe/H] $\ltsim-1$ in our model, which is a bit larger than $0.37$ in C04 and other observations (Fig.\ref{fig:si}).

For S, because of the hardness of observation, the plateau value has not been established.
Some observations \citep{isr01S,tak02} suggest a sharp increase in [S/Fe] with decreasing [Fe/H] like UV OH result, where SI (6) lines at $\sim 8694 \AA$ are used.
Takada-Hidai et al. (2002) adopted a NLTE model and actually commented that the very high [S/Fe] at [Fe/H] $\sim -2.1$ could be smaller with different temperature, and the S trend is uncertain.
Other recent observations \citep{nis04,tak05} using SI (1) lines at $\sim 9200 \AA$ with LTE models provided plateau values of $\sim 0.3$ and $0.46$.
These are consistent with our prediction of [S/Fe] $=0.37-0.50$ at $-3\ltsim$ [Fe/H] $\ltsim-1$ (Fig.\ref{fig:s}).

Ca is a well observed element, and our model succeeds in reproducing the observed plateau [Ca/Fe] $\sim0.27-0.39$, which is larger by $\sim 0.2$ dex than N97 model (Fig.\ref{fig:ca}).
However, [Ti/Fe] is $\sim 0.4$ dex underabundant overall, which cannot be improved by changing our parameters (Fig.\ref{fig:ti}).
A possible model that enhances Ti is a jet-like explosion with high entropy \citep{mae03}.

\subsubsection{Sodium, Aluminum, and Copper}

As shown in Figure \ref{fig:yield-z}, the abundances of the odd-Z elements show a strong metallicity dependence.
The odd-Z elements are enhanced by the surplus of neutrons in $^{22}$Ne, and $^{22}$Ne is transformed from $^{14}$N by the CNO cycle during He-burning. Thus, smaller amounts of CNO elements result in smaller amounts of the odd-Z elements.
With our one-zone model, the decreasing trend of [(Na,Al,Cu)/Fe] toward lower [Fe/H] is seen more weakly because of the mass dependence, and the resultant trends are in excellent agreement with the observations.
Observationally, the NLTE effect for Na and Al is large for metal-poor stars, and the observational data at [Fe/H] $\ltsim -2$ \citep{mcw95,rya96,cay04,hon04} in Figures \ref{fig:na} and \ref{fig:al} are shifted by a constant of $-0.2$ and $+0.5$, respectively \citep{fre05,asp05}.
McWilliam et al. (1995)'s data show significant offset with higher [Al/Fe], of which reason is unclear \citep{rya96}.

\subsubsection{Potassium, Scandium, and Vanadium}

K, Sc, and V yields are overall much underabundant by $\sim1$ dex compared with observations (Figs.\ref{fig:k}-\ref{fig:v}).
For K, the NLTE effect is corrected by a constant shift of $-0.35$ \citep{cay04}.
To solve these problems, UN05 have introduced a low-density model, where the density is assumed to be reduced during explosive burning. This model enhances the $\alpha$-rich freeze-out and thus the Sc production.
This is based on the idea that a relatively weak jet expands the interior of the progenitor before a strong jet forms a strong shock to explode the star.
Alternatively, \citet{fro06} showed that the delayed neutrino mechanism that leads to $Y_{\rm e}>0.5$ in the innermost region gives larger production of Sc, Ti and Zn.
\citet{yos06} have added neutrino processes to explosive nucleosynthesis of UN05, and found larger Sc, V, and Mn production by a factor of ten.
Such additional physical processes would actually work even for zero metallicity stars.

\subsubsection{Chromium, Manganese, Cobalt, and Nickel}

McWilliam et al. (1995) and Ryan et al. (1996) found the decreasing trend of [(Cr,Mn)/Fe] and the increasing trend of [Co/Fe] toward lower metallicity.
These trends have been first explained by Nakamura et al. (1999) by the stellar mass-dependent mass-cut, because Co is synthesized by complete Si-burning in the deepest layer of the ejecta while Cr and Mn form in the outer incomplete Si-burning layers.
Then, UN05 found that a larger explosion energy enhances the amount of Co and decreases Cr and Mn, and showed that these trends can be explained by the energy effect i) if the interstellar medium is not mixed so that the EMP stars are enriched only by a single supernova \citep{aud95} and ii) if the hydrogen mass swept by supernovae ejecta is proportional to the explosion energy and the metallicity well correlates with the energy.
However, C04 claimed that no relation is found in [Mn/Fe] after 0.4 dex correction for low metallicity.

In our chemical evolution model (Figs.\ref{fig:cr}-\ref{fig:ni}), since we use the yields for typical SNe II and HNe, no such trends are seen.
Instead, we focus on whether our mean values meet the observations at [Fe/H] $\sim-2.5$.
As a whole, our yields are in better agreement with observations than N97 yields, although $\sim 0.1$ dex offsets still remain: $\sim +0.2, -0.1, -0.1$, and $-0.1$ dex offsets for Cr, Mn, Co, and Ni, respectively.
However, these elements are much affected by the mixing-fallback process, the explosion energy, and the electron excess $Y_{\rm e}$, so that it would be possible to find a better set by fine-tuning of the parameters under our model.
We also note that there is an inconsistency of observed Cr abundances taken with Cr I and Cr II lines, and higher value of Cr II is favored for our model.

From [Fe/H] $\sim-1$, SNe Ia start to contribute, which results in the increasing [Mn/Fe] toward higher metallicity.
This is confirmed both observationally and theoretically, and Mn can be a key element to discuss the SN Ia contribution, HN fraction, and IMF.
Cr and Co are produced not only from SNe Ia but also from SNe II with [Cr,Co/Fe] $\sim0$ at [Fe/H] $\gtsim-1$.
The observed [Ni/Fe] is $\sim0$ for all range of [Fe/H], and the Ni overabundance of SNe Ia can be reduced: Ni yield depends on $Y_{\rm e}$ in the burning region, which is determined by electron capture and thus sensitive to the propagation speed of burning front and the central density of the white dwarf \citep{nom84,iwa99}.

\subsubsection{Zinc}

The most important agreement of our yields lies in Zn, which is an important element to discuss the cosmic chemical enrichment and is observed in the damped Ly$\alpha$ systems without the dust depletion effect.
The production of the isotopes of Zn depends on metallicity.
At higher metallicity ($Z\ge0.004$), $^{66}$Zn and $^{68}$Zn are synthesized by the neutron capture processes during He and C burning.
At lower metallicity, only $^{64}$Zn is formed in the deep complete Si-burning region and needs to be mixed out into the ejecta in order for enough amount to be ejected.
$^{64}$Zn can be enhanced only by the high energy explosions of HNe (UN03, UN05).
At $-2.5\ltsim$ [Fe/H] $\ltsim-1.5$, [Zn/Fe] is constant $\sim 0.1$, mildly increases to $\sim0.2$ from [Fe/H] $\sim-1.5$ due to the metallicity effect, and mildly decreases to $\sim0$ from [Fe/H] $\sim-1$ due to SNe Ia, which are all consistent with observations (Fig.\ref{fig:zn}).

Primas et al. (2000) found the increasing [Zn/Fe] trend toward lower metallicity at [Fe/H] $\ltsim-3$.
UN05 explained this by the energy effect as well as other trends of iron-peak elements.
Such a trend is not realized in our chemical evolution model.
The contribution of pair-instability supernovae, which produce much more Fe and less [Zn/Fe], is not observed in EMP stars.
Recently, Ohkubo et al. (2006) proposed that core-collapse very massive stars with $M\sim 500-1000M_\odot$ also produce large [Zn/Fe].

\subsubsection{Carbon and Nitrogen}

Since the enrichment from stellar winds and the contribution of low and intermediate mass stars are not included in our chemical evolution model, C and N are overall underabundant than observed.
The major sources of C and N are AGB stars and/or Wolf-Rayet (WR) stars. 
Intermediate mass stars ($3-8M_\odot$) may not be favored because no strong increase is observed in [C/Fe] from [Fe/H] $\sim-2$ to $\sim-1$ \citep{pra94}.
The decrease from [Fe/H] $\sim -1$ in our model is due to the SN Ia contribution. Because no such decrease is observed, low-mass stars with $M<2M_\odot$ may also contribute to C and N production.

With our yields, [C/Fe] and [N/Fe] are $\sim0.4$ dex smaller than observations at [Fe/H] $\sim0$ but larger than N97 yields (Figs.\ref{fig:c} and \ref{fig:n}).
The ejected $^{12}$C mass is comparable to N97 yields, and the increase of [C/Fe] is caused by the difference in the Fe mass (i.e., the larger and smaller Fe mass than N97 are adopted for massive and low-mass SNe II, respectively).
The ejected $^{14}$N mass is ten times larger than N97 yields, because a larger amount of convective mixing of H into the He burning layer takes place in our progenitor stars.
The N production in metal-free stars strongly depends on the detail treatment of the convective mixing, and can be increased by more than a factor of ten \citep{iwa05}.

In the observational data, there are many stars that show large enhancement of carbon abundances ([C/Fe] $>>0$), and some of them also shows a large enhancement of Mg.
As already mentioned, [Zn/Fe] can put constraint on the enrichment source.
With our yield set, massive SNe II can increase [C/Fe] to $\sim 0.4$,  but [Zn/Fe] is much smaller than 0.
HNe can increase [Zn/Fe], but [C/Fe] cannot be larger than 0.
The stars with [C/Fe] $>0.4$ suggest other enrichment sources, which could be i) a single supernova \citep{ume03,iwa05}, namely, a faint supernova with $E_{51}<1$), ii) a few supernovae \citep{lim03b}, or iii) external enrichment from a binary companion \citep[e.g.,][]{sud04}.

\subsubsection{Other elements}

Figure \ref{fig:xfe} shows the [X/Fe]-[Fe/H] diagrams also for the other elements.
Neon and Argon are $\alpha$-elements with plateau values of [Ne/Fe] $\sim 0.5$ and [Ar/Fe] $\sim 0.3$ at [Fe/H] $\ltsim -1$, and can be tested with X-ray observations.
Phosphorus, Chlorine, and Argon can be seen in the damped Ly-$\alpha$ systems.
Fluorine is an important element to examine the neutrino process.

\section{Chemical Evolution of Halo, Bulge, and Thick Disk}
\label{sec:halo}

\subsection{Galactic Halo Model} 

Here we apply our yields to chemical evolution models for the bulge and halo of the Milky Way Galaxy.
For the halo, the MDF of field stars selected with Hipparcos kinematics shows a peak at low [Fe/H] $\sim-1.6$ \citep{chi98}, and has no metal-rich component that is seen in the MDF of globular clusters \citep{zin85}.
This suggests that the efficiencies of star formation and chemical enrichment are very low in the halo.
We use a closed-box model that allows the outflow of material.
The driving source of the outflow should be the feedback
from supernovae. Thus the outflow rate is assumed to be $R_{\rm out} \equiv \frac{1}{\tau_{\rm o}}f_{\rm g}$,
which is proportional to the SFR ($\psi \equiv \frac{1}{\tau_{\rm s}}f_{\rm g}$, see K00 for the other formulation).
Timescales of $\tau_{\rm s}=8$ Gyr and $\tau_{\rm o}=1$ Gyr are adopted to meet the MDF.
The adopted parameters in our chemical evolution models are summarized in Table \ref{tab:param}.

\begin{figure*}
\center
\includegraphics[width=16.5cm]{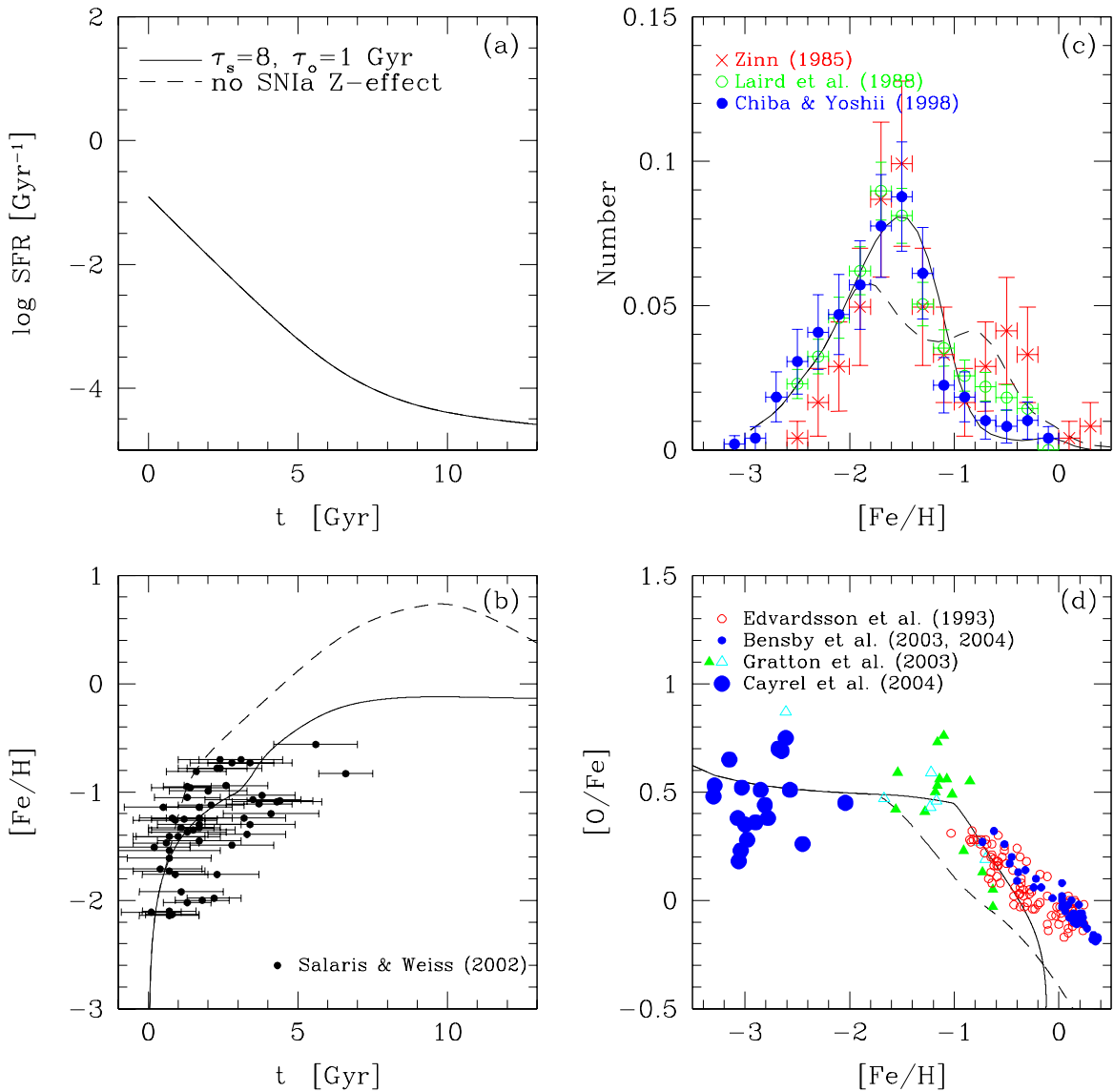}
\caption{\label{fig:halo}
The chemical evolution of the Galactic halo; (a) the star formation rate, (b) the age-metallicity relation, (c) the metallicity distribution function, and (d) the [O/Fe]-[Fe/H] relation.
The solid and short-dashed lines are for the outflow model with and without the SN Ia metallicity effect, respectively.
Observational data sources are:
(b) filled circles, \citet{sal02}; 
(c) crosses, \citet{zin85}; open circles, \citet{lai88}; filled circles, \citet{chi98};
and (d) small open circles, \citet{edv93}; small filled circles, thin disk stars in \citet{ben04}; filled and open triangles, \citet{gra03}; large filled circles, \citet{cay04}.
}
\end{figure*}

Figure \ref{fig:halo} shows (a) the SFR, (b) the age-metallicity relation, (c) the MDF, and (d) the [O/Fe]-[Fe/H] relation.
The solid and dashed lines show the cases with and without the metallicity effect of SNe Ia.
The metallicity increases quickly to [Fe/H] $\sim-1.5$ at $t\sim0.9$ Gyr, and then the outflow becomes effective and chemical enrichment takes place slowly.
If we do not include the SN Ia metallicity effect, [Fe/H] keeps on increasing, which results in the double peaks in the MDF (panel c).
Without the metallicity effect, [$\alpha$/Fe] starts to decrease at [Fe/H] $\sim-1.7$, which is earlier than observed (panel d).
Since our SN Ia progenitor scenario requires the companion mass range of $\sim 1-3M_\odot$, the SN Ia lifetime spans over $0.5-20$ Gyr.
It is difficult to delay the onset of the decrease in [$\alpha$/Fe] without the metallicity effect (K98).

\subsection{Galactic Bulge Model} 

For the bulge, observational results of the MDF are controversial (Fig.\ref{fig:bulge}c).
McWilliam \& Rich (1994) showed a broad MDF that extends to [Fe/H] $\sim 1$.
However, a narrow MDF is shown recently with a sub-solar peak and a sharp cut-off at [Fe/H] $\sim 0$ \citep{zoc03}.
We thus construct several models to meet each observation:

\noindent
(A) For McWilliam \& Rich (1994)'s MDF, a simple infall model ($R_{\rm in} \equiv \frac{1}{\tau_{\rm i}}\exp(-\frac{t}{\tau_{\rm i}})$) with a short star formation timescale ($\tau_{\rm s}=\tau_{\rm i}=1$) gives a good agreement (dashed line).
With our model A, the SFR is peaked at $\sim 1$ Gyr, and the star formation continues until the present, producing a lot of low [$\alpha$/Fe] stars in the bulge.

\noindent
(B) On the other hand, to reproduce Zoccali et al. (2003)'s MDF, star formation needs to be terminated somehow at $t_{\rm w}=3$ Gyr (solid line), possibly by supernova induced galactic winds or by the feedback from the active galactic nuclei.
To meet this MDF, $\tau_{\rm s}=0.5$ and $\tau_{\rm i}=5$ Gyr are adopted.
With our model B, [Fe/H] does not increase after $t\sim3$ Gyr, and [$\alpha$/Fe] cannot be lower than 0.

The short timescale of chemical enrichment is imprinted in the [$\alpha$/Fe]-[Fe/H] diagram (panel d), namely in the decrease of [$\alpha$/Fe] toward higher [Fe/H] than $-1$ \citep{mat90}.
However, McWilliam \& Rich (2004) claimed that the abundance patterns of bulge stars look peculiar, and the evolutions of [O/Fe] and [(Mg,Si)/Fe] are different.
Both our models show the [$\alpha$/Fe] decrease from [Fe/H] $\sim-0.7$, which is consistent with the observed O trends.
However, Mg observation does not show such decrease, which may suggest that the chemical enrichment timescale is much shorter. 
For example, with a flatter IMF model of $x=1.1$, [$\alpha$/Fe] is almost constant until [Fe/H] $\sim-0.3$ (dotted line).
(Since a larger amount of metals are ejected for $x=1.1$ than for Salpeter IMF, the time duration of chemical enrichment should be shorter, i.e., $t_{\rm w} = 2$ Gyr, in order to meet the MDF.)
The resultant MDF is shifted to higher metallicity and the number of metal-poor stars with [Fe/H] $\ltsim-1$ is as small as in the observation by Ramirez et al. (2000).
In these bulge models, the chemical enrichment timescale is so short that [Fe/H] reaches $-1.1$ at $t\sim0.4$ Gyr, and no difference is realized for the cases with and without the SN Ia metallicity effect.

\subsection{Thick Disk}

Several formation scenarios of the Galactic thick disk have been debated for the following observational features:
i) lack of vertical gradients of metallicity,
ii) existence of SN Ia contribution,
iii) larger [$\alpha$/Fe] than in thin disk stars,
iv) older age than the thin disk,
and v) lack of very metal-poor G-dwarfs \citep[e.g., ][]{gil95,fel04}.

\begin{figure*}
\center
\includegraphics[width=16.5cm]{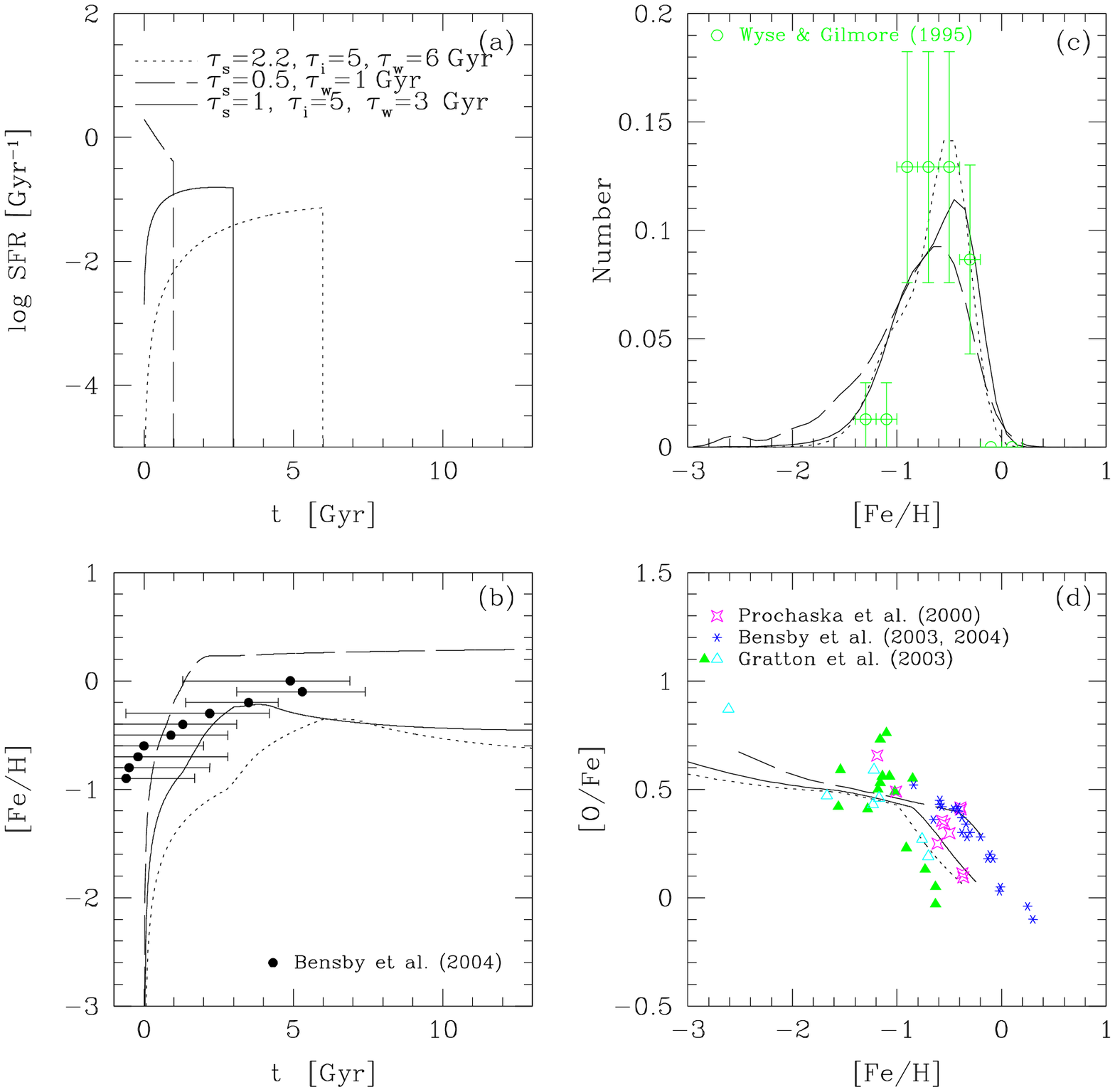}
\caption{\label{fig:thick}
The same as Fig.\ref{fig:halo}, but for the Galactic thick disk.
The dotted line is for the same model as the solar neighborhood, but with a truncated SFR.
The dashed and solid lines are for the closed-box and infall models with short star formation timescales.
Observational data sources are:
(b) filled circles, \citet{ben04b};
(c) open circles, \citet{wys95}; 
and (d) four-pointed stars, \citet{pro00}; filled and open triangles respectively for dissipative and accretion component in \citet{gra03}; asterisks, thick disk stars in \citet{ben04}.
}
\end{figure*}

Here we construct several models for the thick disk as well as the Galactic halo and bulge.
Although we use the one-zone model, we do not assume that the thick disk is formed monolithically, and violent heating and satellite accretion scenarios are not excluded.
Because of the lack of very metal-rich and very young stars in thick disk, we assume that star formation is truncated at $t=t_{\rm w}$ (Fig.\ref{fig:thick}a).
With these SFRs, the available MDF can be reproduced (panel c).
However, the age-metallicity (panel b) and [O/Fe]-[Fe/H] (panel d) relations obtained from these models are different, with which we can put a constraint on the star formation history as follows:

\noindent
(A) If the thick disk is formed as well as the thin disk, we can adopt the same SFR as the solar neighborhood model in \S \ref{sec:chem}, but with the cutoff at $6$ Gyr (dotted line).
It is assumed that the stars formed in the solar neighborhood model at $t > 6$ Gyr and $t \le 6$ correspond to the thin and thick disk stars, respectively.
In this scenario, chemical enrichment takes place slowly in infalling materials.
Our model A predicts that the relations of the age-metallicity and [$\alpha$/Fe]-[Fe/H] are the same as in the thin disk, which seems to be inconsistent with the available observations.

\noindent
(B) The closed-box model (dashed line) is not viable because of the G-dwarf problem: the lack of very metal-poor stars in the MDF.
In our model B, the peak metallicity of the observed MDF requires such a short timescale of star formation as 0.5 Gyr, which results in the strong initial starburst, a rapid increase of [Fe/H], and much larger [$\alpha$/Fe] at [Fe/H] $\gtsim -1$.

\noindent
(C) The infall model with short star formation timescale (solid line) can meet the observed MDF, and still give larger [$\alpha$/Fe] at [Fe/H] $\gtsim -1$ than the thin disk model, which is as large as in \cite{pro00}.
This scenario may be quite possible.
Our model C predicts that the age-metallicity relation is different from the thin disk, as shown in \cite{ben04b}, and that the duration of star formation is as short as $\sim 3$ Gyr.

\subsection{Discussion} 

\begin{figure*}
\center
\includegraphics[width=16.5cm]{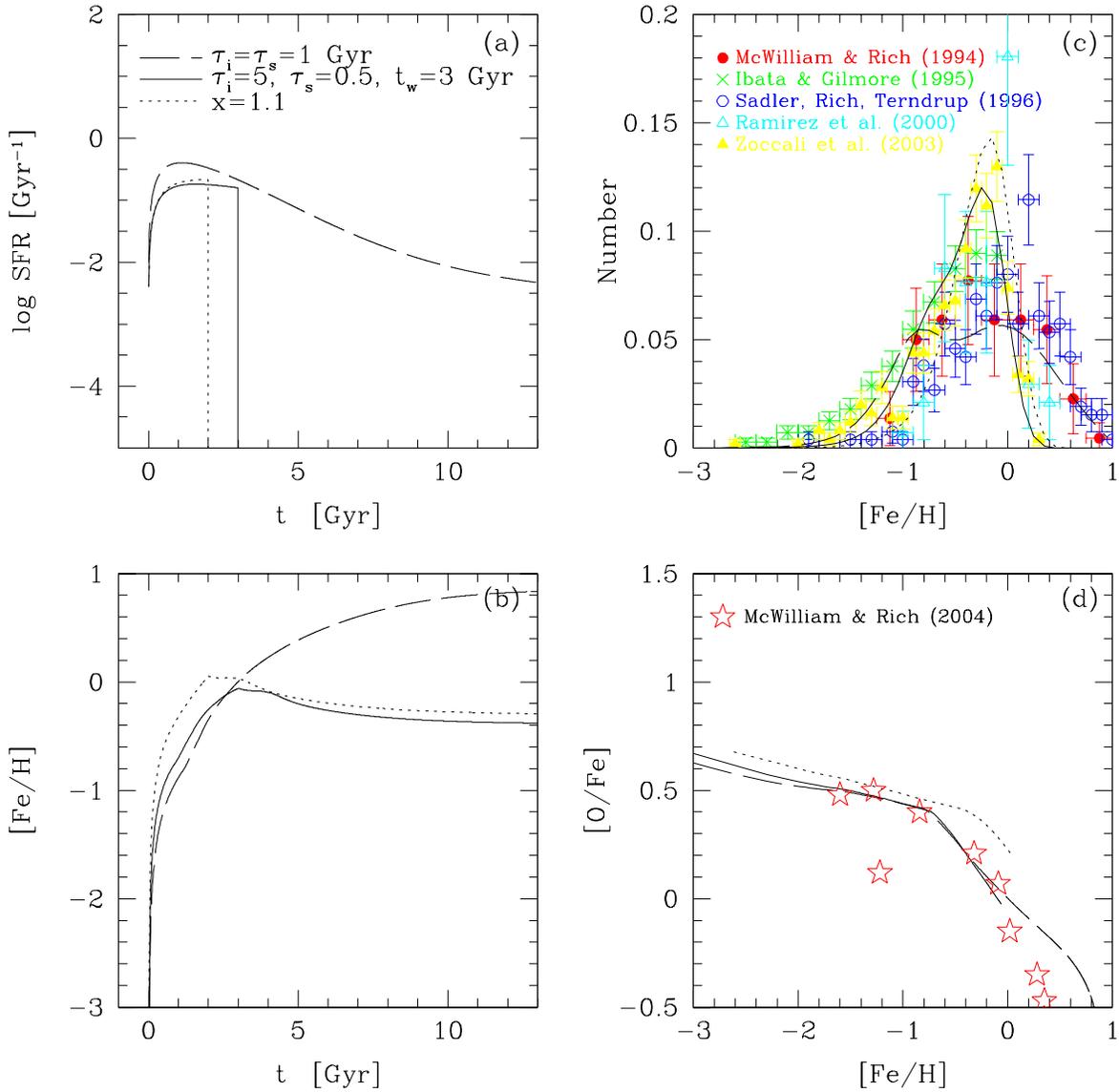}
\caption{\label{fig:bulge}
The same as Fig.\ref{fig:halo}, but for the Galactic bulge.
The dashed and solid lines are for the wind models to give broad and narrow metallicity distribution functions, respectively.
The dotted line is for the model with a flatter IMF to give constant [O/Fe].
Observational data sources are:
(c) filled circles, \citet{mcw94}; crosses, \citet{iba95}; open circles, \citet{sad96}; open triangles, \citet{ram00}; filled triangles, \citet{zoc03}; 
and (d) stars, \citet{mcw04}.
}
\end{figure*}

\begin{figure*}
\center
\includegraphics[width=16.5cm]{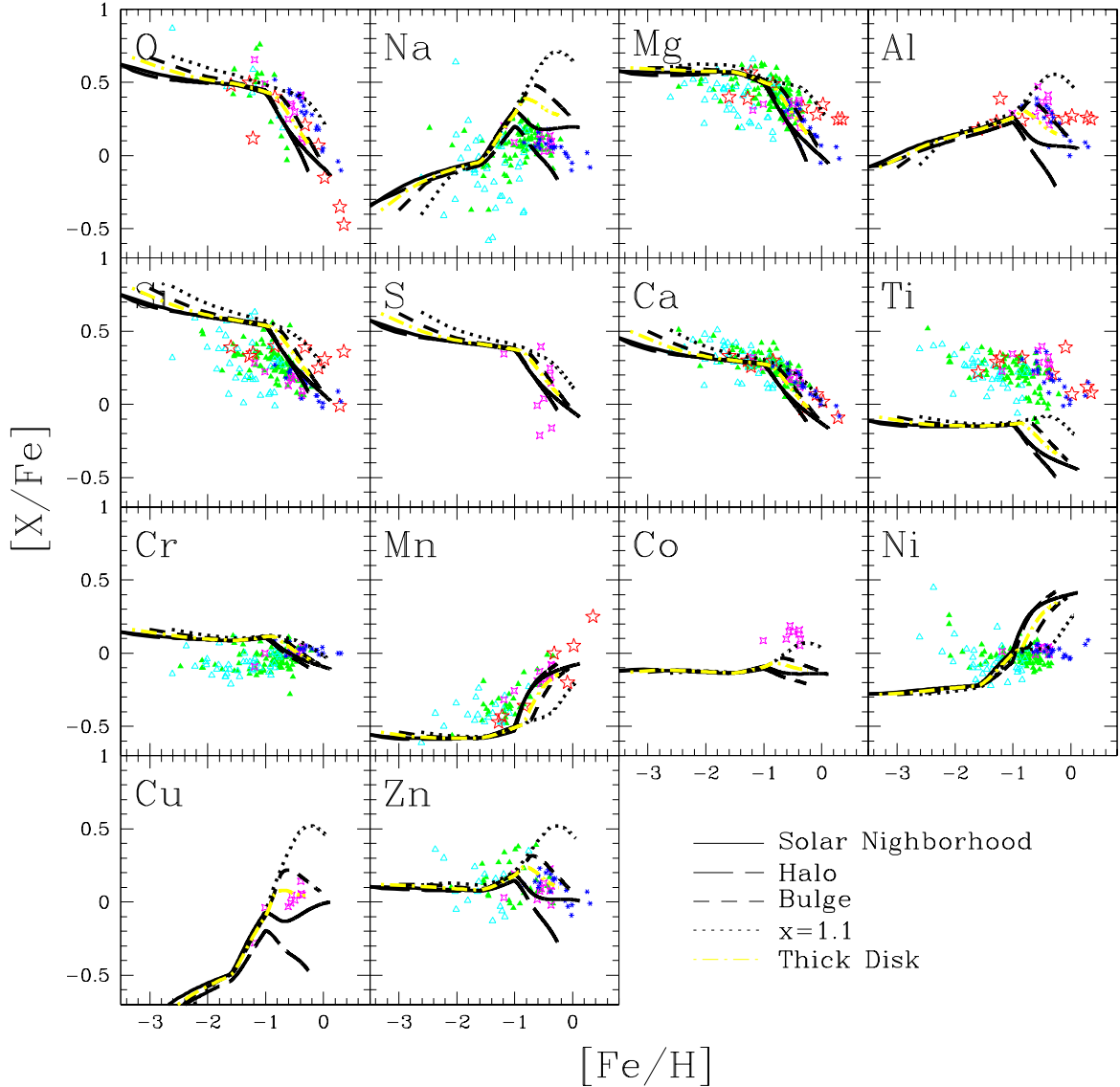}
\caption{\label{fig:xfe2}
[X/Fe]-[Fe/H] relations for the disk (solid line), halo (long-dashed line), and bulge (short dashed line), bulge with a flatter IMF (dotted line), and thick disk (dot-dashed line) models.
Here we take the solar neighborhood model (solid line in Fig.\ref{fig:mdf}) for the disk, the outflow model for the halo (solid line in Fig.\ref{fig:halo}), the bulge models with the Salpeter IMF (solid line in Fig.\ref{fig:bulge}) and the flat IMF (dotted line in Fig.\ref{fig:bulge}), and the infall model for the thick disk (dotted line in Fig.\ref{fig:thick}).
Observational data sources are:
For thick disk stars, four-pointed stars, \citet{pro00}; filled and open triangles respectively for dissipative and accretion component in \citet{gra03}; small asterisks, \citet{ben04}.
For bulge stars, large stars, \citet{mcw04}.
}
\end{figure*}

In Figure \ref{fig:xfe2}, we compare the [X/Fe]-[Fe/H] relations for the Galactic disk (solid line), halo (long-dashed line), bulge (short-dashed line), bulge with a flat IMF (dotted line), and thick disk models (dot-dashed line).
Observational data of thick disk stars are shown \citep{pro00,ben03,gra03}, except for the large stars are for bulge stars \citep{mcw04}.
Here we take the solar neighborhood model (solid line in Fig.\ref{fig:mdf}) for the disk, the outflow model for the halo (solid line in Fig.\ref{fig:halo}), the bulge models (B) with the Salpeter IMF (solid line in Fig.\ref{fig:bulge}) and the flat IMF (dotted line in Fig.\ref{fig:bulge}), and the infall model (C) for the thick disk (solid line in Fig.\ref{fig:thick}).
We note that these models are constructed based on the observations of the limited regions (solar neighborhood and Baade's window), and the radial dependencies on star formation and chemical enrichment histories are neglected.
Since, in all models, the same stellar yields are adopted and the initial metallicity is set to be primordial, differences among model predictions are due to differences in the SFR, the IMF, and the SN Ia contribution.

Since the metallicity effect on SNe Ia is included for all models, all models give similar results at [Fe/H] $\ltsim -1$ where SNe Ia do not contribute.
The plateau values of [$\alpha$/Fe] at $-2 \ltsim$ [Fe/H] $\ltsim -1$ depend on the IMF, because [$\alpha$/Fe] is larger for larger stellar mass.
The flat IMF model gives 0.1 dex larger [$\alpha$/Fe] than Salpeter IMF.
The increasing trends of [$\alpha$/Fe] and decreasing trends of [(Na,Al,Cu)/Fe] toward lower metallicity are, respectively, originated from the mass and metallicity dependences.
The slope of these trends depends on the SFRs, namely, the chemical enrichment timescale, which is short in our bulge models.
Therefore, the slope is steep in our bulge models.

From [Fe/H] $\sim -1$, [$\alpha$/Fe] decreases because of the SN Ia contribution.
While the decreasing point of [$\alpha$/Fe] is simply determined by the SN Ia metallicity effect in the disk and halo, but it is determined by the SN Ia lifetime in the bulge where the chemical enrichment timescale is short enough.
In such systems as bulge, we can safely discuss the formation timescale from the [$\alpha$/Fe] decreasing point.
For faster chemical enrichment, [$\alpha$/Fe] starts decreasing at larger [Fe/H].
In our bulge model with Salpeter IMF, [$\alpha$/Fe] decreases from [Fe/H] $\sim -0.7$, while from [Fe/H] $\sim -0.4$ with the flat IMF model.

Among $\alpha$-elements, McWilliam \& Rich (2004) showed different trends for the bulge (stars).
The observed O trend is well reproduced with the Salpeter IMF model, while the flat IMF model is favored from the constant Mg/Fe.
As noted before, it is difficult to explain the different evolution of O and Mg with our models.
Some uninvolved physics such as strong stellar winds might be important.
For Si, observational data shows a large scatter, and is consistent with both models.
S and Ca can be produced also by SNe Ia to some extent, and thus the observed Ca trend is consistent with both models.

At the same [Fe/H], iron-peak abundance ratios also change, and [Mn/Fe] ratio increases quickly because more Mn is produced by SNe Ia than Fe, relative to solar abundance (i.e., [Mn/Fe] $>0$).
The odd-Z abundance ratios [(Al,Na,Cu)/Fe] increase to be super-solar with a peak at [Fe/H] $\sim-0.4$, because of the yield metallicity dependence and the SNe Ia contribution.
[Zn/Fe] may put a constraint on the IMF because Zn yield depends strongly on the mass and Zn is mainly produced from HNe and metal-rich massive SNe.
With the flat IMF model, larger enhancement of [Zn/Fe] is predicted at [Fe/H] $\gtsim -1$ than Salpeter IMF.
This does not depend on our assumption of the constant hypernova efficiency $\epsilon_{\rm HN}$ because massive SNe produce Zn as much as HNe at $Z \ge 0.004$.

The formation timescale of the system can be constrained from the elemental abundance ratios and the metallicity distribution function.
However, since observational results are still controversial, we summarize our predictions focusing on a part of the observational results.
If there is not many super-solar metal stars in the bulge like Zoccali et al. (2003)'s MDF, star formation should be truncated by somehow, and the duration of star formation should be $\sim 3$ Gyr.
Even if no such truncation is included, the star formation timescale should be as short as $\sim 1$ Gyr, and most stars are as old as $\sim 10$ Gyr.
On the other hand, the decrease in [(O,Ca)/Fe] and the increase in [Mn/Fe] by SNe Ia require that star formation continues longer than $\sim 1$ Gyr.
The flat [(Mg,Si)/Fe] may suggest that the IMF is flatter than Salpeter IMF and the chemical enrichment timescale is much shorter.
In this case, the duration of star formation should be shorter ($\sim 2$ Gyr) than the Salpeter IMF case, in order to reproduce the same MDF.
It is important to confirm the different trends of O and Mg with large sample, and Zn observation is also interesting.

For thick disk stars, similar observational feature is seen.
Prochaska et al. (2000) showed the abundance patterns of thick disk stars (four-pointed stars). They found that [(O,Si,Ca)/Fe] decrease with increasing [Fe/H], while [(Mg,Ti)/Fe] are constant, which are similar to the bulge stars except for Si.
Bensby et al. (2004) also showed that the abundance patterns of thin (circles) and thick (asterisks) disk are different, and [(O,Mg,Al)/Fe] and [Zn/Fe] of thick disk are larger than thin disk.
In the thick disk models constructed to meet the narrow MDF by \cite{wys95}, star formation and chemical enrichment take place slightly faster than in our bulge model.
The resultant [X/Fe]-[Fe/H] relations are similar to those of the bulge model, but with slightly earlier decrease of [$\alpha$/Fe] from [Fe/H] $\sim -0.8$, which is roughly consistent with observations (small points except for large stars).
Therefore, we would conclude that the thick disk is as old as the bulge, and the formation timescale is as short as $\sim 1-3$ Gyr.

\section{Conclusions}

We calculate the evolution of heavy element abundances from C to Zn in the solar neighborhood adopting our new nucleosynthesis yields.
Our new yields are based on the new developments in the observational/theoretical studies of supernovae and extremely metal-poor stars in the Galactic halo.
We use the light curve and spectra fitting of individual supernova to estimate the mass of the progenitor, explosion energy, and produced $^{56}$Ni mass.

\begin{figure*}
\center
\includegraphics[width=16cm]{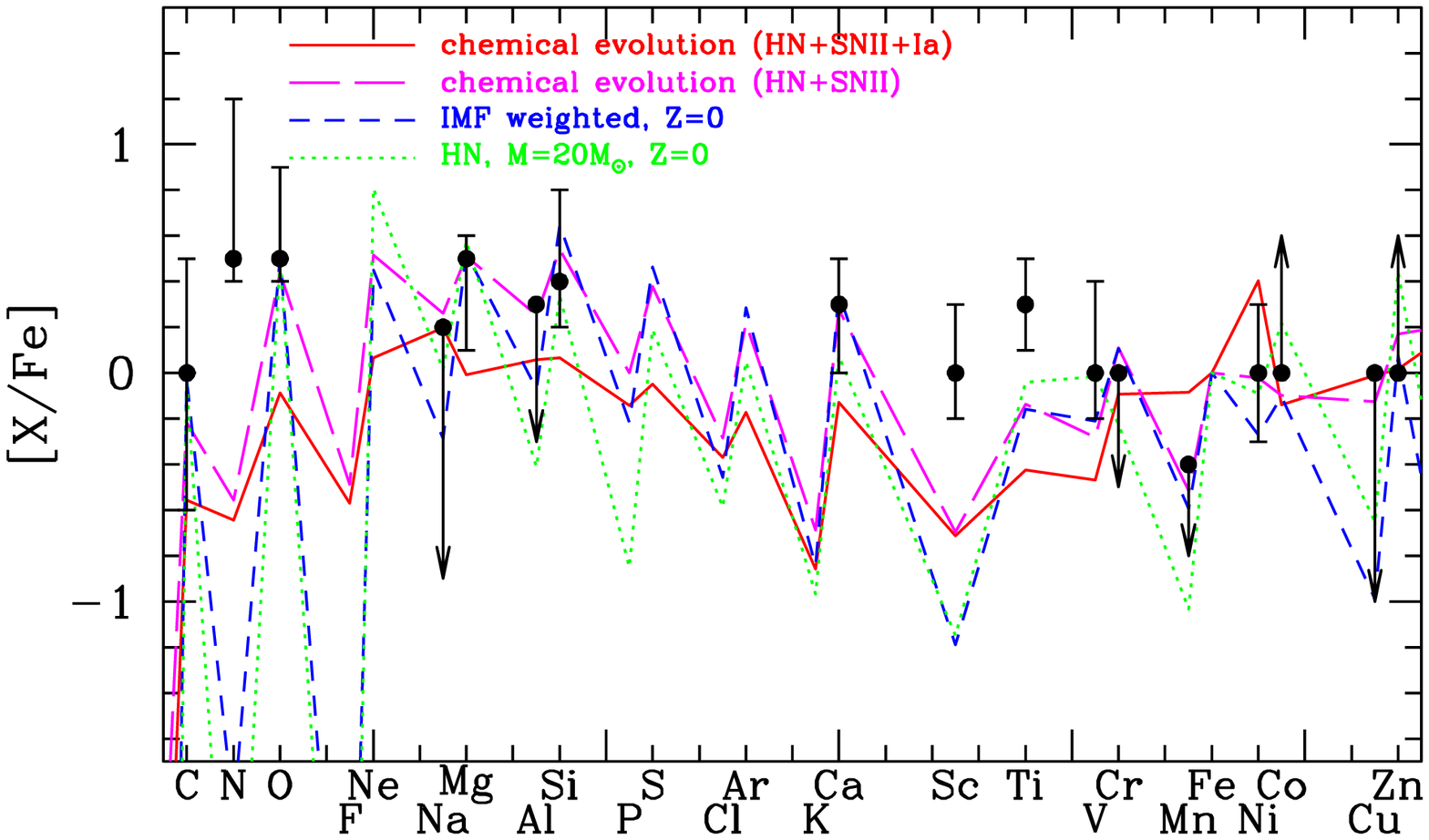}
\vspace*{-3cm}
\caption{\label{fig:imf}
The solid and long-dashed lines show the abundance patterns with our chemical evolution model at [Fe/H] $=0$ and $-1.1$, respectively, which correspond to the solar abundance [X/Fe] $=0$ and the IMF weighted SN II yield without SN Ia contribution, respectively.
The dots, errorbars, arrows show the observations for the plateau value at $-1.5 \ltsim$ [Fe/H] $\ltsim -1$ \citep{sne91,mel02,gra03}, the scatter at $-3.5 \ltsim$ [Fe/H] $\ltsim -2.5$, and the trend toward [Fe/H] $\sim -4$ \citep{mcw95,rya96,cay04,hon04}, respectively.
The short-dashed and dotted lines show the IMF-weighted yield with $Z=0$ and the HN yield with $M=20M_\odot, E_{51}=10, Z=0$, respectively.
}
\end{figure*}

The elemental abundance ratios are in good agreement with observations.
Figure \ref{fig:imf} provides a summary of our new yield table set.
The solid and long-dashed lines show the abundance patterns in our chemical evolution model for the solar neighborhood at [Fe/H] $=0$ and $-1.1$, respectively, which should correspond to the solar abundance [X/Fe] $=0$ and the IMF weighted SN II yield without SN Ia contribution, respectively.
The metallicity and energy dependencies of the yields are demonstrated by the short-dashed and dotted lines, which show the IMF-weighted yield with $Z=0$ and the HN yield with $M=20M_\odot, E_{51}=10, Z=0$, respectively.

\begin{enumerate}
\item
The solar abundance (i.e., [X/Fe] $=0$) is basically well reproduced with our chemical evolution model at [Fe/H] $=0$ (solid line).
Some of the discrepancies between our model and observations may be explained as follows:

\noindent
(1) 
The underabundances of C and N may suggest that AGB stars and Wolf-Rayet star winds are the dominant sources of these elements.

\noindent
(2) 
F, K, Sc, Ti, and V are underabundant, which cannot be increased by changing our parameters such as metallicity and energy. 
A jet-like explosion \citep{mae03} can efficiently increase Ti and Zn abundances and a low-density model (UN05) or neutrino process \citep{fro06,wan06,yos06} may increase Sc abundance.

\noindent
(3)
Ni overproduction by SNe Ia can be reduced \citep{iwa99}.

\item
The observed plateau values at $-1.5 \ltsim$ [Fe/H] $\ltsim -1$ \citep[dots, ][]{sne91,mel02,gra03} are in good agreement with our model at [Fe/H] $=-1.1$ (long-dashed line), where SNe Ia do not contribute.
Only Ti, Sc and V are underabundant by 0.4, 0.6 and 0.2 dex, respectively, relative to Fe.

\item
The observed scatter at $-3.5 \ltsim$ [Fe/H] $\ltsim -2.5$ (errorbars) and the observed trends toward [Fe/H] $\sim -4$ \citep[arrows,  ][]{mcw95,rya96,cay04,hon04}
may be due to {\sl inhomogeneous} enrichment, and could be explained with the variations of the properties of individual supernovae such as energy and metallicity (UN05).

\item
The metallicity effect is seen in our IMF-weighted yield of metal-free stars (short-dashed line), where the abundance ratios of the odd-Z elements [(Na, Al, Mn, Cu, ...)/Fe] are smaller than those of the even-Z elements by $\sim 0.6$ dex.
The observed trends (arrows) are well reproduced with this metallicity effect in our chemical evolution model.

\item
The energy effect is seen in our HN yield for $M=20M_\odot, E_{51}=10$, and $Z=0$ (dotted line), where the abundance ratios of iron-peak elements [(Cr, Mn, Co, and Zn)/Fe] are different from normal SNe II.
The observed trends (arrows) could be explained with the variation of the explosion energy of individual supernovae.

\item

[$\alpha$/Fe] (O, Mg, Si, S, Ca, and Ti) depends on the mass of the progenitor star and is larger for massive SNe II.
For HNe, however, because the iron yield is as large as the yields of $\alpha$-elements, [$\alpha$/Fe] is almost constant.
This may account for the observed small scatter of [$\alpha$/Fe] in the solar neighborhood, being independent of the mixing process of interstellar medium.
The small [$\alpha$/Fe] (O, Mg, Si, S, Ca, and Ti) in some anomalous stars is due to i) SNe Ia, ii) relatively large Fe production from low-mass SNe II with $M=13-15M_\odot$, iii) large Fe production from HNe, or iv) the SN I.5 explosion of some AGB stars (see \S \ref{sec:xfe}).
The large [$\alpha$/Fe] is due to i) small Fe production from SNe II with $E_{51} \le 1$, or ii) large fallback mass for HNe.
From Zn and iron-peak elements, we can distinguish the enrichment source of such stars;
HNe produce large [(Zn,Co)/Fe], and SNe Ia and I.5 produce relatively large [Mn/Fe].

\end{enumerate}

\acknowledgments
This work has been supported in part by the Grant-in-Aid for Scientific Research (17030005, 17033002, 18104003, 18540231) and the 21st Century COE Program (QUEST) from  the Japan Society for Promotion of Science (JSPS) and the Ministry of Education, Culture, Sports, Science, and Technology (MEXT) of Japan.
C.K., N.T., and T.O. thank to the JSPS for a financial support. 
We would like to thank K. Maeda and M. Shirouzu for fruitful discussion.

\clearpage
\newpage

\clearpage

\clearpage

\clearpage

\newpage
\clearpage
\begin{deluxetable}{lccccccc}
\tablecaption{\label{tabsn}
The nucleosynthesis yields for the Type II Supernovae in the ejecta in $M_\odot$.}
\tablewidth{0pt}
\footnotesize
\tablehead{
$M$       &       13&        15&        18&        20&        25&        30&        40\\
$E_{51}$       &        1&         1&         1&         1&         1&         1&         1
}
\startdata
\multicolumn{8}{c}{$Z=            0$}\\
\hline
$M_{\rm final}$ &13.00 &15.00 &18.00 &20.00 &25.00 &30.00 &40.00\\
$M_{\rm cut}$ &      1.57&      1.48&      1.65&      1.66&      1.92&      2.07&      2.89\\
p         &  6.60E+00&  7.58E+00&  8.43E+00&  8.77E+00&  1.06E+01&  1.17E+01&  1.40E+01\\
d         &  1.49E-16&  1.69E-16&  1.28E-16&  8.66E-17&  2.02E-16&  1.34E-16&  3.46E-16\\
$^3$He    &  4.12E-05&  4.09E-05&  3.33E-05&  4.76E-05&  2.11E-04&  2.06E-04&  2.57E-05\\
$^4$He    &  4.01E+00&  4.40E+00&  5.42E+00&  5.94E+00&  8.03E+00&  9.52E+00&  1.19E+01\\
$^6$Li    &  3.65E-23&  1.11E-22&  4.37E-23&  3.65E-21&  2.69E-21&  1.13E-22&  7.54E-22\\
\enddata
\end{deluxetable}
\begin{deluxetable}{lcccc}
\tablecaption{\label{tabhn}
The nucleosynthesis yields for Hypernovae in the ejecta in $M_\odot$.}
\tablewidth{0pt}
\footnotesize
\tablehead{
$M$       &       20&        25&        30&        40 \\
$E_{51}$       &       10&        10&        20&        30 
}
\startdata
\multicolumn{5}{c}{$Z=         0$}\\
\hline
$M_{\rm final}$ &20.00 &25.00 &30.00 &40.00\\
$M_{\rm cut}$ &      1.88 &2.80 &3.27 &5.53\\
p         &  8.77E+00&  1.06E+01&  1.17E+01&  1.40E+01\\
d         &  8.66E-17&  2.06E-16&  1.09E-14&  1.66E-14\\
$^3$He    &  4.76E-05&  2.11E-04&  2.06E-04&  2.57E-05\\
$^4$He    &  5.96E+00&  8.03E+00&  9.54E+00&  1.18E+01\\
$^6$Li    &  1.53E-22&  3.22E-20&  3.50E-17&  5.39E-17\\
\enddata
\end{deluxetable}

\begin{deluxetable}{lccccc}
\tablecaption{\label{tab2}
The IMF weighted yields with $x=1.35, M_{\ell}=0.07M_\odot, M_{\rm u}=50M_\odot$.}
\tablewidth{0pt}
\small
\tablehead{
	  &\multicolumn{4}{c}{SNII+HN}   &SNIa      \\
$Z$       &         0&    0.001&     0.004&      0.02&          
}
\startdata
p         &  3.28E-02&  3.14E-02&  2.96E-02&  2.45E-02&  0.00E+00\\
d         &  5.76E-18&  2.21E-15&  1.97E-16&  5.34E-16&  0.00E+00\\
$^3$He    &  2.51E-07&  5.24E-07&  6.52E-07&  7.06E-07&  0.00E+00\\
$^4$He    &  2.30E-02&  2.26E-02&  2.22E-02&  2.05E-02&  5.66E-03\\
$^6$Li    &  1.70E-20&  1.52E-19&  6.49E-19&  1.76E-18&  0.00E+00\\
$^7$Li    &  8.67E-12&  6.67E-13&  3.14E-15&  5.85E-14&  0.00E+00\\
$^9$Be    &  1.25E-20&  7.74E-19&  6.54E-19&  7.26E-19&  0.00E+00\\
$^{10}$B  &  1.86E-17&  2.38E-14&  1.47E-13&  4.25E-13&  0.00E+00\\
\enddata
\end{deluxetable}

\newpage

\begin{deluxetable}{l|ccccc}
\footnotesize
\tablewidth{0pt}
\tablecaption{\label{tab:param}
Parameters of chemical evolution models; Infall, star formation, and outflow timescales, and the galactic wind epoch in Gyr. $x$ denotes the slope of the initial mass function.}
\tablehead{
 & $\tau_{\rm i}$ & $\tau_{\rm s}$ & $\tau_{\rm o}$ & $\tau_{\rm w}$ & $x$
}
\startdata
solar neighborhood           & 5 & 2.2 & - & - & 1.35 \\
halo                         & - & 8   & 1 & - & 1.35 \\
bulge, model (A)              & 1 & 1   & - & - & 1.35 \\
bulge, wind model (B)           & 5 & 0.5 & - & 3 & 1.35 \\
bulge, flat IMF model        & 5 & 0.5 & - & 2 & 1.10 \\
thick disk, disk-like model (A)  & 5 & 2.2 & - & 6 & 1.35 \\
thick disk, closed-box model (B) & - & 0.5 & - & - & 1.35 \\
thick disk, infall model (C)    & 5 & 1   & - & 3 & 1.35 \\
\enddata
\end{deluxetable}

\end{document}